\documentclass[12pt,article]{iopart}
\def\bm#1{\mbox{\boldmath $#1$}}   

\usepackage{graphicx}
\usepackage{amssymb}
\usepackage{amsmath}
\usepackage{comment}
\usepackage{float}	   	 
\usepackage{appendix}
\begin{document}

\title{Quantum gates in coupled quantum dots controlled by coupling modulation}
\author{Alejandro D. Bendersky, Sergio S. Gomez, Rodolfo H. Romero}
\address{Instituto de Modelado e Innovación tecnológica, CONICET-UNNE,  Facultad de Ciencias Exactas y Naturales y Agrimensura, Universidad Nacional del Nordeste\\ Avenida Libertad 5500 (3400) Corrientes, Argentina }


\date{\today}
\begin{abstract}
We studied the dynamics of a pair of single-electron double quantum dots (DQD) under longitudinal and transverse static magnetic fields and time-dependent harmonic modulation of their interaction couplings. We propose to modulate the tunnel coupling between the QDs to produce one-qubit gates and the exchange coupling between DQDs to generate entangling gates, the set of operations required for quantum computing. We developed analytical approximations to set the conditions to control the qubits and applied them to numerical calculations to test the accuracy and robustness of the analytical model. The results shows that the unitary evolution of the two-electron state performs the designed operations even under conditions shifted from the ideal ones.
\end{abstract}





\section{Introduction}
%
Electron spins in semiconductor quantum dots are envisioned as good prospects for future quantum computing platforms as well as interesting physical systems for the study of novel quantum phenomena in the nanoscale \cite{Loss98,Kloeffel13,Watson18,Zhang19,Liu19,Xue22,Ferraro20}. 
Electrostatically defined quantum dots are fabricated by application of gate voltages to electrodes deposited upon semiconductor heterostructures allowing to control the number of electrons contained, their spatial extension and their energy spectra \cite{Hanson07, Zwanenburg13, Zwerver22}.
Additional external static and variable electric and magnetic fields allows to manipulate both the spatial and spin degrees of freedom and their interactions \cite{Petta05, Nowack07}, giving rise to a wide variety of qubit proposals and performances \cite{Chatterjee21, Stano22, Burkard23}. \\

The electric control of electron spins in semiconductor quantum dots can be achieved using electric dipole spin resonance (EDSR) \cite{Koppens06, Tokura06, Pioro08}. The spin resonance is obtained by periodically displacing the electrons around their respective equilibrium positions in a slanting field. Electron movement is controlled by plunger and gate electrodes and the transverse field is produced by a micromagnet giving an effective synthetic spin-orbit (SO) interaction. \cite{Pioro08, Zhang21}. Control of qubits are used to generate single and two-qubit gates, that is, unitary operations acting on the states of the physical system used as quantum computing platforms \cite{Burkard23}. 
Algorithms for quantum computation can be expressed in terms of a set of elementary single- and two-qubits quantum gates, e. g., rotations, phase shift and CNOT, what allows for the universal representation of arbitrary unitary transformations \cite{Makhlin02, Zajac18}



A number of qubits based on different electron number and configurations have been advocated for implementation \cite{Burkard23}. 
A single electron in a double quantum dot with controllable interdot tunnel coupling and detuning can be driven to localize or delocalize its wave function between both QDs. 
The presence of an external uniform longitudinal magnetic field and a transverse magnetic gradient allows one to use electric spin dipole resonance (EDSR) to manipulate the electron spin by electrical means. 
This mode of operation, dubbed as flopping-mode qubit \cite{Benito19, Teske23}, has been investigated in Si-MOS QD \cite{Kawakami14, Hu23} and Ge holes \cite{Mutter21, Hendrickx20, Watzinger18, Zhang25}. 
Furthermore, two-qubit operations requires the physical interaction between two qubits. Exchange coupling interaction has been used for implementing two-qubit quantum gates in various type of semiconductor spin qubits \cite{Li12, Klinovaja12, Wardrop14, Martins16, Yang20, Nguyen23, Zhou24}

Hybridization of spatially delocalized bonding and antibonding states, due to the hopping between the QDs, and Zeeman levels for spin projection parallel and antiparallel, due to the longitudinal magnetic field, gives a four-level spectrum. It comprises the highest and lowest widely separated levels having a energy gap increasing with the longitudinal field strength, and two central close-lying levels that becomes degenerate when the tunneling coupling is a half of the longitudinal Zeeman splitting. The transverse inhomogeneous magnetic field breaks the degeneracy and introduces a small gap between these two levels. 

In this work, we use this quasidegenerate pair of levels to encode the qubits and exchange interaction to couple them. Time-dependent modulation of magnetic fields or level detuning have been proved useful to control single-electron flopping mode qubits \cite{Hajati24, Gyorgy22}. Here, we produce single-qubit and two-qubit entangling gates by harmonic modulation of the tunnel coupling around the working point, and by biharmonic modulation of the exchange interaction between a pair of double quantum dots, at specific frequencies, respectively. 
The resulting quantum gates operate within intervals shorter than coherence time, and have the accuracy usually required for the application of quantum error computing codes.

The structure of the paper is as follows. 
In Section \ref{model} we present the model of the system and the definition of the computational and leakage spaces. 
In Section \ref{1-qubit gates} we study analytically the unitary evolution of the states of a single DQD under a time-dependent harmonic tunneling, to derive conditions defining one-qubit gates. The accuracy of the analytical approximations are assessed by comparing them to numerical simulations of the exact model as measured from the magnitude of the leakage out from the computational space as well as the infidelity. 
Section \ref{2-qubit gates} defines the two-qubit gates from the approximate analytical unitary evolution of time-dependent exchange-coupled DQDs and test their accuracy and robustness through numerical calculations. 
Finally, Section \ref{conclusions} provides a summary of the results and some concluding remarks.

\section{Model \label{model}}
The model consists of two coupled single-electron double quantum dots (DQDs), described by the Hamiltonian 
\begin{equation}
    H= H_0^{(1)} + H_0^{(2)} + H_{\rm exch}, 
    \label{system-Hamiltonian}
\end{equation}
where $H_0^{(i)}$ ($i=1, 2$) is the Hamiltonian for the single electron in the $i$-th DQD and $H_{\rm exch}=J\hbar {\bf s}_1\cdot{\bf s}_2$ is the exchange interaction between them.

\begin{figure}
    \centering
    $\begin{array}{cccc}
${\large (a)}$ & \includegraphics[scale=0.3]{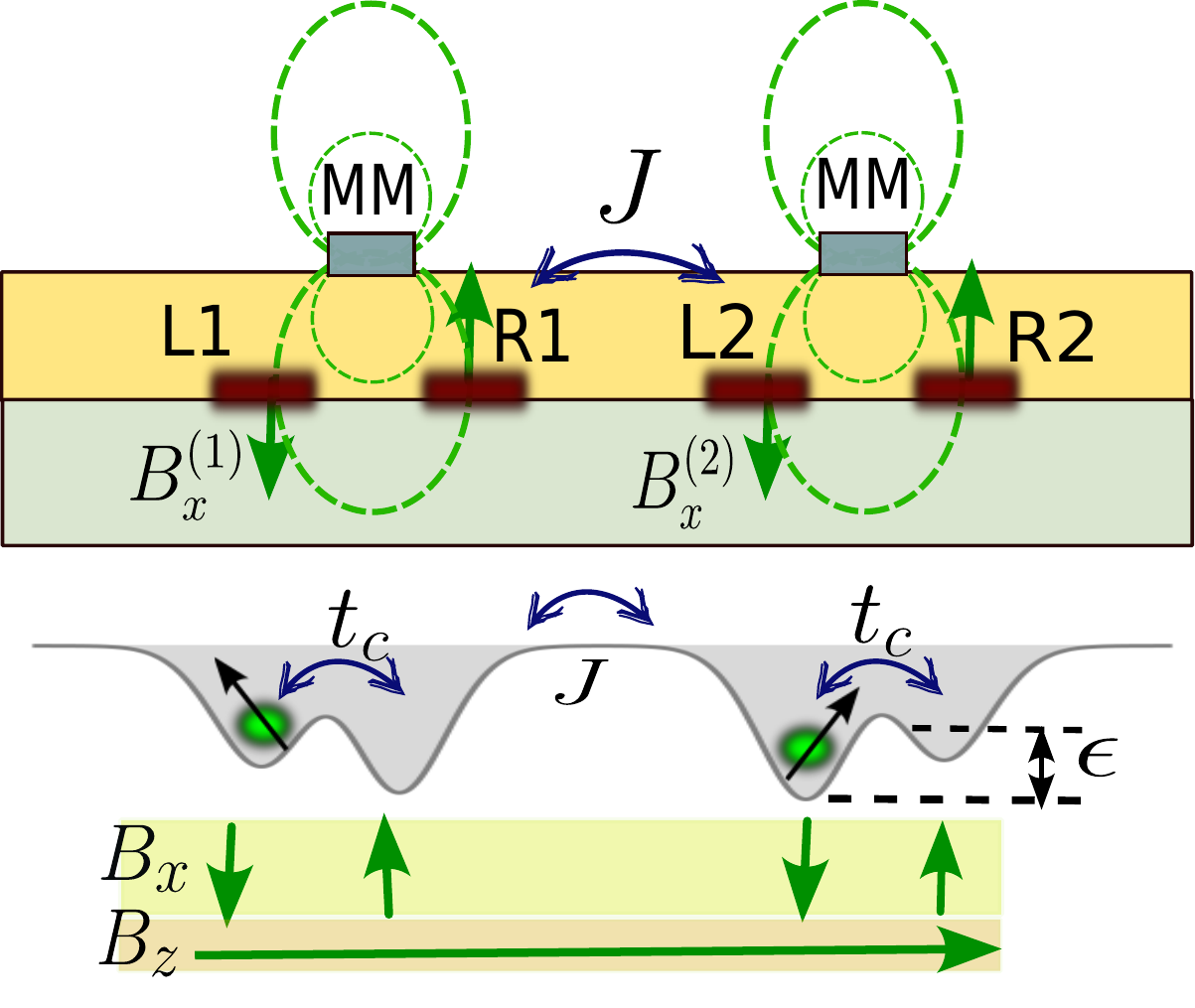} & 
${\large (b)}$ & \includegraphics[scale=0.3]{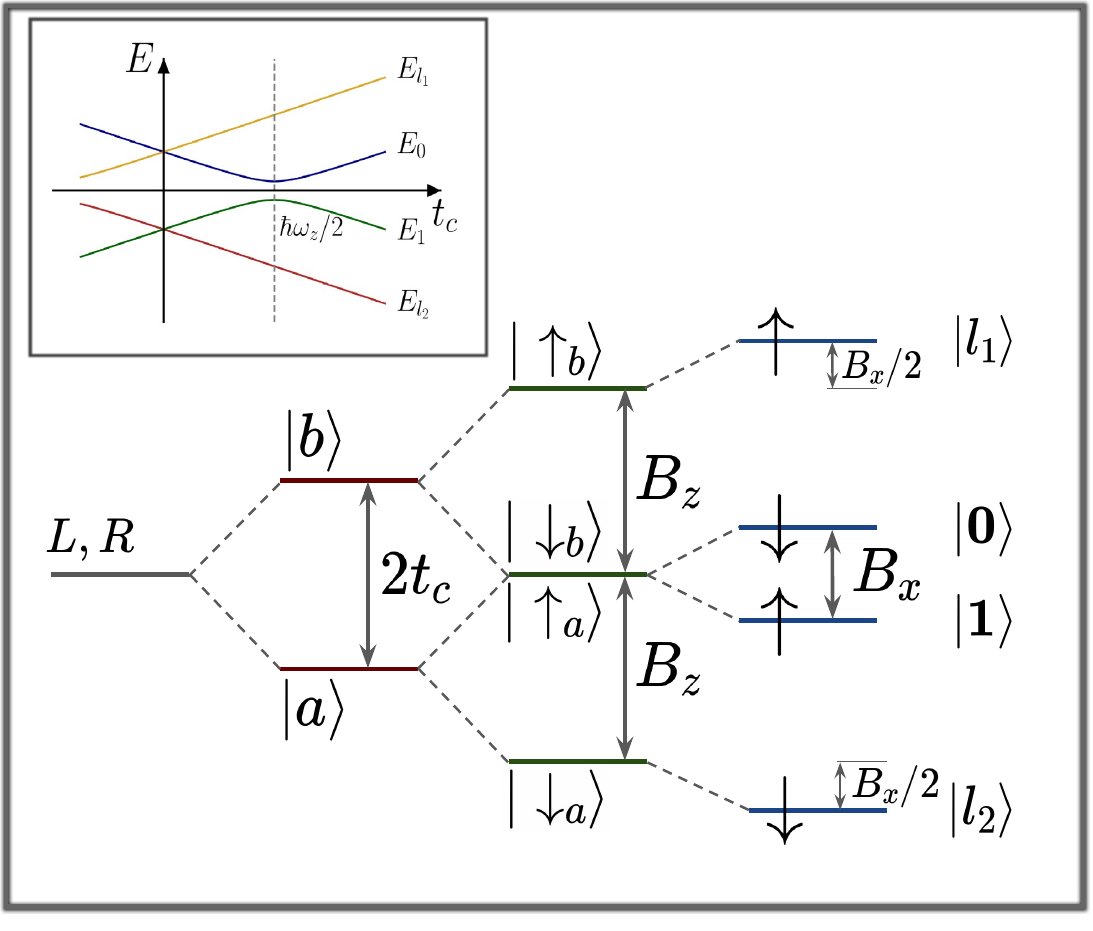}
    \end{array}$
    \caption{Two exchange-coupled single-electron flopping-mode qubits. (a) Scheme of the cross section of the device and its physical model in terms of electrostatically-defined potential wells and barriers. Double QDs (L1, R1) and (L2,R2) are subject to longitudinal and transverse magnetic fields produced by the micromagnets MM. 
(b) Scheme of energy levels of each single-electron DQD in absence of exchange coupling. From left to right the various mechanism involved in the setup: the coupling $t_c$ shifts bonding and antibonding states, the longitudinal magnetic field $B_z$ breaks spin degeneracy and fix the quantization axis and, finally, the inhomogeneous component $B_x$ hybridize $|\downarrow_b\rangle$ and $|\uparrow_a\rangle$. Inset: energy labels as a function of $t_c$ showing quasi-degeneration at $t_c=g\mu_B B_z/2$. }
    \label{fig:model}
\end{figure}

Each electrically defined DQD contains a single electron whose energy levels and tunel coupling $t_c$ are controlled by their detuning $\epsilon$ --the energy shift between their ground states--, and by the hight and width of the potential barrier between the left (L) and right (R) QDs, as sketched in Figure \ref{fig:model}a.  The exchange interaction can be electrostatically controlled by the barrier between the double QDs. We assume the system is set in the (1, 1) charge state, i.e., with each DQD being singly occupied.

A homogeneous magnetic field $B_z$ along the interdot segment introduces Zeeman energy shifts $\pm \hbar\omega_z/2 =\pm g\mu_B B_z/2$ for $\uparrow$ and $\downarrow$ spin states, and a nearby micromagnet induces a transverse inhomogeneous magnetic field $B_x$ with opposite values $\pm B_x$ at QD$_{\rm L}$ and QD$_{\rm R}$ and contributes to a Zeeman energy $\pm \hbar\omega_x/2= \pm g\mu_B B_x/2$. This transverse magnetic field gradient plays the role of an artificial spin-orbit interaction allowing transitions between states with different $s_z$-projections. We assume a longitudinal gradient $\partial B_x/\partial z$, such that the transverse fields $B_x^{(1)}$ and $B_x^{(2)}$ are different at each DQD, such that any of them is typically one order of magnitude smaller than $B_z$.



\subsection{Single-electron DQD eigenstates}
The Hamiltonian for each single-electron DQD in the basis of spatially localized spin states $\{|\uparrow\rangle,|\downarrow\rangle\}\otimes\{|L\rangle,|R\rangle\}$ is $H_0=(\epsilon \tilde{\tau}_z + \hbar\omega_z \sigma_z + 2t_c \tilde{\tau}_x + \hbar\omega_x \sigma_x \tilde{\tau}_z)/2$, where the $\sigma_{x,z}$ and $\tilde{\tau}_{x,z}$ are, respectively, the spin and space Pauli matrices in this basis. Transforming to antibonding $(-)$ and bonding $(+)$ basis $\{|a\rangle,|b\rangle\}=\{(|L\rangle \pm|R\rangle)/\sqrt{2}\}$, the Pauli matrices for the space degrees of freedom are transformed as $(\tilde{\tau}_{x},\tilde{\tau}_{z})\rightarrow ({\tau}_{z},{\tau}_{x})$, i.e. 
\begin{equation}
    H_0 = \frac{1}{2}(\epsilon \tau_x + \hbar\omega_{z} \sigma_z + 2t_c \tau_z + \hbar\omega_x  \sigma_x \tau_x).
\end{equation}
The transformation transfers the opposite signs of $B_x$ in QD$_{\rm L}$ and QD$_{\rm R}$, to a $\hbar\omega_x$ shift contributing to bonding ($+$) or antibonding ($-$) states in them.  The term $\epsilon \tau_x$ couples the subspaces ${\cal H^-}$, spanned by $\{|\uparrow_a\rangle,|\downarrow_b\rangle\}$, and ${\cal H^+}$, spanned by $\{|\uparrow_b\rangle,|\downarrow_a\rangle\}$. 
In this basis, $\{|\uparrow_a\rangle,|\downarrow_b\rangle,|\uparrow_b\rangle,|\downarrow_a\rangle\}$, the Hamiltonian becomes
\begin{equation}
H_0  =  \frac{1}{2} \left(
  \begin{array}{cccc}
      \hbar\omega_z-2t_c & \hbar\omega_x & \epsilon & 0  \\
      \hbar\omega_x & -\hbar\omega_z+2t_c & 0 & \epsilon \\
      \epsilon & 0 & \hbar\omega_z+2t_c & \hbar\omega_x \\
      0 & \epsilon & \hbar\omega_x & -\hbar\omega_z-2t_c  
  \end{array}
  \right)
\end{equation}
 A variable time-dependent detuning $\epsilon(t)$ would allow to control the electric dipole moment of the DQD and, therefore, its coupling to the field of a resonant cavity \cite{Benito19}. 
Here, however, we propose to operate the DQD at zero bias permitting to decouple ${\cal H}^-$ and ${\cal H}^+$. 
The Hamiltonian for zero detuning becomes block-diagonal.

Figure \ref{fig:model}b shows the energy levels of the double QD at zero bias; while the bonding ($b$) and antibonding ($a$) levels are separated $2t_c$ by the tunnel coupling, the field $B_z$ shifts spins $\uparrow$ and $\downarrow$ upwards and downwards, respectively. By setting $t_c=\hbar\omega_z/2$, the states $|\uparrow_a\rangle$ and $|\downarrow_b\rangle$ becomes degenerate, while the lowest ($|\downarrow_a\rangle$) and highest ($|\uparrow_b\rangle$) ones becomes separated by $2\hbar\omega_z$. The inhomogeneous transverse field $B_x$ breaks the degeneracy of the central pair of states.

At zero detuning the electron becomes spatially delocalized between QD$_{\rm L}$ and QD$_{\rm R}$ and oscillates coherently between them. Due to the  transverse field $B_x$ takes opposite values at both QDs, the electron experiences a variable slanting field causing transitions between opposite spin states \cite{Tokura06, Pioro08}. 

\subsection{Computational and leakage spaces}
We assume the system to be prepared at the working point $t_c=\hbar\omega_z/2$, such that $H_0$ becomes block diagonalized as $H_0=H_c \oplus H_l$.

The computational space es spanned by the eigenstates of 
$H_c=\sigma_x\hbar\omega_x/2$, 
\begin{equation}
|0\rangle,|1\rangle= (|\uparrow_a\rangle\pm |\downarrow_b\rangle)/\sqrt{2},     
\label{eq:comp-states}
\end{equation}
with $E_{0,1}=\hbar\omega_{0,1}= \pm\hbar\omega_x/2$. 
On the other hand,  $H_l = (\sigma_z\cos\theta+\sigma_x\sin\theta)\hbar\omega/2$ have eigenstates of $H_l$
\begin{eqnarray}
    |l_1\rangle &=& \sin(\theta/2) |\uparrow_b\rangle + \cos(\theta/2) |\downarrow_a\rangle \nonumber\\
    |l_2\rangle &=& \cos(\theta/2) |\uparrow_b\rangle - \sin(\theta/2) |\downarrow_a\rangle, 
    \label{eq:leakage-states}
\end{eqnarray}
where $\sin\theta=\omega_x/\omega$, with eigenenergies 
\begin{equation}
E_{l_1}=-E_{l_2}= \hbar\omega=\hbar \sqrt{\omega_x^2+4\omega_z^2}.     
\label{eq:leakage-energies}
\end{equation}
becomes leakage levels due to detuning ($\epsilon\ne 0$) in each QD, or due to the exchange interaction ($J\ne0$) between both electrons in the two double QDs. 
For large $B_z$, $|l_1\rangle$ approaches $|\downarrow_a\rangle$, and $|l_2\rangle$ approaches $|\uparrow_b\rangle$.

The inset of Figure \ref{fig:model}b shows that this working point corresponds to a minimum of the gap $E_g=\hbar(\omega_1-\omega_0)$ in the qubit space, so that it becomes protected against tunnel coupling noise throughout first order ($\partial E_g/\partial t_c =0$). \\

Therefore, in the basis $\{|0\rangle,|1\rangle,|l_1\rangle,|l_2\rangle\}$, the one-electron Hamiltonian is $H_0=(\hbar/2){\rm diag}(\omega_x,-\omega_x, \omega, -\omega)$, such that $E_{l_1}> E_0 > E_1 > E_{l_2}$. 
The partition into computational and leakage spaces, ${\cal H}_{\rm comp}=\{|0\rangle, |1\rangle\}$ and ${\cal H}_{\rm leak} =\{|l_1\rangle, |l_2\rangle\}$, allows one to write the two-electron Hilbert space in block form as 
\begin{equation}
    {\cal H} = ({\cal H}_{\rm comp}\oplus {\cal H}_{\rm leak})^{\otimes 2},
    \label{complete-Hilbert}
\end{equation}
where ${\cal H}_{\rm comp}^{\otimes 2}$ is where we shall prepare the qubits and define the two-qubit operations; ${\cal H}_{\rm leak}^{\otimes 2}$ are two-electron product states fully orthogonal to the qubit space and only accessible from the computational space via the exchange interaction. 

Then, in the absence of exchange interaction ($J=0$), the two-electron energy spectrum of ${\cal H}_{\rm comp}^{\otimes 2}=\{|00\rangle,|11\rangle,|01\rangle,|10\rangle\}$ is the sum and the difference of the single electron transverse Zeeman energies. That is,  the non-interacting two-electron Hamiltonian will be $H_0^{(1)}+H_0^{(2)} = (\hbar/2) {\rm diag} (\Omega_1,-\Omega_1,\Omega_2,-\Omega_2)$ where $\Omega_1=\omega_x^{(1)}+\omega_x^{(2)}$, and $\Omega_2=\omega_x^{(1)}-\omega_x^{(2)}$. 
The energy eigenvalues $\pm\hbar\Omega_1/2$ correspond to non-interacting electrons with either both at higher or both at lower single-electron state, i.e., $|00\rangle$ or $|11\rangle$. Similarly, $\pm\hbar\Omega_2/2$ correspond to one electron at the higher and the other at lower level, i.e., states $|01\rangle$ or $|10\rangle$.
Note that the assumption of different transverse magnetic fields at each DQD breaks the degeneracy that otherwise would exist at $\Omega_2=0$. 
Similarly,  ${\cal H}_{\rm leak}^{\otimes 2}=\{|l_1l_2\rangle,|l_2l_1\rangle,|l_1l_1\rangle,|l_2l_2\rangle\}$ have energy eigenvalues $\pm(E_{l_1}+E_{l_2}) = \hbar\Delta\Omega$ and $2E_{l_1}=-2E_{l_2}\approx 4 \hbar\omega_z$.
The frequency $\Delta\Omega$ corresponds to that of the energy of the state $|l_1l_2\rangle$, Eq. (\ref{eq:leakage-energies}), where one electron is in the higher positive $E_{l_1}=\hbar(\omega_x^{(1)2}+4\omega_z^2)^{1/2}$, while the other is in the lower negative energy $E_{l_2}=-\hbar(\omega_x^{(2)2}+4\omega_z^2)^{1/2}$ leakage states of each double QD. Therefore, up to order ${\cal O}(\omega_x/\omega_z)$, 
\begin{equation}
    \Delta\Omega = \frac{\omega_x^{(2)2}-\omega_x^{(1)2}}{2\omega_z}=\frac{\Omega_1\Omega_2}{2\omega_z}.
    \label{eq:DeltaOmega}
\end{equation}
\subsection{Exchange interacting Hamiltonian}
Let us consider now the effect of switching on the electron-electron interaction by opening the tunneling between the L2 and R1 quantum dots (Figure \ref{fig:model}a). The presence of the exchange interaction $H_{\rm exch}$ allows for transitions ${\cal H}_{\rm comp} \leftrightarrow {\cal H}_{\rm leak}$ between states from the computational and leakage spaces, and $H$ is a $8\times 8$ matrix in the complete ${\cal H}^{\otimes 2}_{\rm comp} \oplus {\cal H}^{\otimes 2}_{\rm leak}$ Hilbert space, Ec. (\ref{complete-Hilbert}), whose dynamics can be calculated numerically. 

Transforming to the spin-adapted leakage basis $\{|\chi_n\rangle\}\ (n=1,\ldots, 4)$, where $|\chi_{1,2}\rangle=(|l_1l_2\rangle\pm |l_2l_1\rangle)/\sqrt{2}$ and $|\chi_{3,4}\rangle=(|l_1l_1\rangle\pm |l_2l_2\rangle)/\sqrt{2}$, the transition amplitudes between qubit states and the highly excited states $|\chi_3\rangle$ and $|\chi_4\rangle$, have contributions ${\cal O}(J\Omega_i/\omega_z)\ll J$, and are neglected here. 
Then, we work with the effective approximate Hamiltonian in the basis $\{|00\rangle, |11\rangle,|\chi_{1}\rangle, |\chi_{2}\rangle, |01\rangle, |10\rangle \}$  
\begin{equation} \label{eq:hamiltonian-2qb}
    H = \frac{\hbar}{4}\left( 
    \begin{array}{cccccc}
        2\Omega_1 & J & \sqrt{2}J & 0 & 0 & 0 \\
        J & -2\Omega_1 & -\sqrt{2}J & 0 & 0 & 0 \\ 
        \sqrt{2}J & -\sqrt{2}J & -J & 4\Delta\Omega & 0 & 0  \\
        0 & 0 & 4\Delta\Omega & -J & \sqrt{2}J & -\sqrt{2}J \\
        0 & 0 & 0 & \sqrt{2}J & 2\Omega_2 & J \\
        0 & 0 & 0 & -\sqrt{2}J & J & -2\Omega_2
    \end{array}
    \right).
\end{equation}
We shall perform one-qubit operations by modulating the tunnel coupling $t_c(t)$, and two-qubit operations by modulating the exchange coupling $J(t)$. Furthermore, $\Delta\Omega\ll\Omega_1,\Omega_2$ [see Eq. (\ref{eq:DeltaOmega})] only induces first-order transitions between the leakage states $|\chi_1\rangle$ and $|\chi_2\rangle$, but not from them to the qubit states. Therefore, we shall define the quantum logical operations from the simpler block-diagonal Hamiltonian ($\Delta\Omega=0$)
\begin{equation}
    H = \frac{\hbar}{4} \left(
    \begin{array}{ccc}
          2\Omega_1 & J & \sqrt{2}J \\
        J & -2\Omega_1 & -\sqrt{2}J \\
        \sqrt{2}J & -\sqrt{2}J & -J 
    \end{array}
    \right) \oplus
    \left(
    \begin{array}{ccc}
       -J & \sqrt{2}J & -\sqrt{2}J \\
         \sqrt{2}J & 2\Omega_2 & J \\
         -\sqrt{2}J & J & -2\Omega_2
    \end{array}
    \right),
    \label{H6x6}
\end{equation}
where $J$ induces transitions within each three-state block at the same time.  
\section{Single-qubit gates \label{1-qubit gates}}
\subsection{Unitary evolution from tunneling modulation}
When the exchange interaction between the DQDs is switched off ($J=0$), each qubit becomes independent. We introduce single-qubit operations by adding a time-dependent modulation to the coupling around the working point, i.e., $t_c(t)=t_c+\delta t_c(t) = \hbar \omega_z/2 + \eta_{c} \sin\omega t$, with amplitude $\eta_{c}$ and frequency $\omega$. Then, the single-qubit space takes the form of a harmonically driven two-level system
\begin{equation}
    H_{\rm TLS} = \frac{1}{2}\left(
\begin{array}{cc}
    \hbar\omega_x & 2\delta t_{c}(t) \\
    2\delta t_{c}(t) & -\hbar\omega_x
\end{array}
    \right).
\end{equation}
Within the rotating wave approximation (RWA), Rabi oscillations with frequency $\Omega=\sqrt{\Delta^2+(\eta_c/\hbar)^2}$ are obtained, where $\Delta= \omega-\omega_x$ is the frequency detuning of the excitation $\omega$ with respect to the energy gap $\hbar\omega_x=g\mu_B B_x$.

The corresponding evolution operator, within this approximation, is 
    \begin{eqnarray}
  \fl  U_1(\eta_c,\Delta,t) &=& \left( 
    \begin{array}{cc}
       \left( \cos\frac{\Omega t}{2} +i\frac{\Delta}{\Omega}\sin\frac{\Omega t}{2}\right) e^{-i(\omega_x+\Delta) t/2} & 
       (\eta_c/\Omega) \sin\frac{\Omega t}{2} e^{-i(\omega_x+\Delta)  t/2} \\
       & \\
     -(\eta_c/\Omega) \sin\frac{\Omega t}{2} e^{i(\omega_x+\Delta)  t/2} & 
     \left( \cos\frac{\Omega t}{2} -i\frac{\Delta}{\Omega}\sin\frac{\Omega t}{2}\right) e^{i(\omega_x+\Delta) t/2}
    \end{array}
    \right)
    \label{U_RWA}
\end{eqnarray}
such that $U_1(0,\Delta,t) = {\cal R}_z(\omega_x t)$ becomes a $z$-axis rotation when the modulation is switched off ($\eta_c=0$). On the other hand, by switching on the tunneling modulation $\eta_c$ at resonance ($\Delta=0$), we get
\begin{eqnarray}
    U_{1, \rm res}(\phi,\vartheta) &=& \left( 
    \begin{array}{cc}
        e^{-i \phi/2} \cos\vartheta/2 & e^{-i \phi/2}\sin\vartheta/2  \\
        -e^{i \phi/2}\sin\vartheta/2 &  e^{i \phi/2} \cos\vartheta/2
    \end{array}
    \right) \nonumber \\
    &=& R_z(\phi)R_y(-\vartheta),
    \label{U_RWA_resonant}
\end{eqnarray}
where $\vartheta = \eta_c t/\hbar$ is the phase of the Rabi oscillations at $t$, and $\phi = \omega_x t$. As shown in Table \ref{tab:1-qubit gates}, by properly choosing the amplitude $\eta_c$ and evolution time $T_{\rm gate}$, various other single-qubit gates for quantum computations are obtained; 
$z$-rotations are controlled by the frequency of the gap $\omega_x$, and $y$-rotations by the amplitude of the harmonic modulation of the hopping $\eta_c$. Note that other possibilities can also be given for non-resonant driving.
\begin{table}[htbp]
    \centering
    \begin{tabular}{cccc}
\hline\hline
      $\phi= \omega_x T_{\rm gate}$ & $\vartheta$ & $T_{\rm gate}=\vartheta\hbar/\eta_c$ & Gate  \\ 
      \hline 
      any value $\varphi$ & $2\pi$ &  $2\pi\hbar/\eta_c$ & {\rm Phase}($\varphi$) \\
       $n\pi$ ($n$ odd) & $\pi$  & $\pi\hbar/\eta_c$ & $\hat{X}$ \\
      $n\pi$ ($n$ even) & $\pi$  & $\pi\hbar/\eta_c$ & $\hat{Y}$ \\
      $n\pi$ ($n$ odd) & $2\pi$  & $\pi\hbar/\eta_c$ & $\hat{Z}$ \\
      $n\pi$ ($n$ odd) & $\pi/2$ & $\pi\hbar/2\eta_c$ & $\hat{H}$ \\
    \hline\hline
    \end{tabular}
    \caption{Single qubit quantum gates generated by the resonant RWA evolution operator $U_{1,\rm res} (\phi,\vartheta)$, eq. (\ref{U_RWA_resonant}), with amplitude $\eta_c$ evolving during a time $T_{\rm gate}$.}
    \label{tab:1-qubit gates}
\end{table}
\subsection{Numerical results: Infidelity and leakage}
The evolution of any qubit under the unitary gates defined above, has a negligible leakage as far as (i) the detuning $\epsilon$ vanishes, (ii) the tunneling coupling is set at the particular value $\hbar\omega_z/2$, and (iii) the harmonic modulation is resonant with the frequencies $\omega_x^{(1)}$ or $\omega_x^{(2)}$. Any departure from those requirements would produce unwanted leakage to states out from the computational space.  We show that the errors it introduces in the operations are small for realistic values of the parameters.

In the following we present the results of numerical simulations of the dynamics of the states under the exact Hamiltonian, when the analytical conditions defining the single-qubit gates are not perfectly fulfilled. 
As a measure of departure from the ideally expected conditions, we shall use the average infidelity $I$ and leakage ${\cal L}$ of the operations, both calculated as $1-F$, with \cite{Pedersen07}
\begin{equation}
    F = [{\rm Tr}(MM^\dagger)+|{\rm Tr}(M)|^2]/n(n+1),
    \label{eq:def_Fidelity}
\end{equation}
where $M$ is $M_I=U_0^\dag PU_{\rm num}P$ for the infidelity and $M_{\cal L}=PU_{\rm num}^\dag U_{\rm num}P$ for the leakage; $n$ the dimension of the unitary matrix, $P$ is the projector on the computational space and $U_{\rm num}$ is the numerically calculated unitary evolution within the complete 16-dimensional Hilbert space. With the definitions above, $M_I$ measures the approximation of the computational space-projected unitary evolution $PU_{\rm num}P$ to the ideal gate $U_0$. $M_{\cal L}$ is a measure of the unitarity of $PU_{\rm num}P$ within the embedded computational space, that it should fulfill if it were completely isolated from the rest of the spectrum.\\

We take the longitudinal field fixed at $B_z=0.6$ T ($\hbar\omega_z=34.72 \ \mu$eV), and therefore, the working point is set at the QD coupling $t_c=17.36$ $\mu$eV. We choose a transverse $B_x= 150$ mT ($\omega_x/2\pi = 2.1$ GHz) at one DQD to produce the single-qubit gates. 

Firstly, we analyze the performance of the one-qubit gates under finite detuning $\epsilon$ and shifts from resonance $\Delta\omega_x$ with respect to the analytically designed ones. 
Following the prescriptions from the analytical approximations (Table \ref{tab:1-qubit gates}), applying a harmonic modulation with amplitude $\eta_c=0.643\ \mu$eV at the frequency $\omega_x$ around $t_c$ should generate $H$ and $Z$ gates. That is, $H=U_{\rm num}(\omega_x T_H, \vartheta_H=\pi/2)$ and $Z=U_{\rm num}(\omega_x T_Z,\vartheta_Z=2\pi)$, where the operation times are $T_H=\vartheta_H\hbar/\eta_c=1.6$ ns and  $T_Z=\vartheta_Z\hbar/\eta_c=6.4$ ns. 
In both cases, $\omega_x T_{\rm op}=n_{\rm odd}\pi$, where $n_{\rm odd}$ is an arbitrary odd integer. That is, the parameter $\hbar\omega_x/\eta_c=n_{\rm odd}\pi/\vartheta$ must take the values $2n_{\rm odd}$ for $H$, and $n_{\rm odd}/2$ for $Z$. 

\begin{figure}[H]
    \centering
    \includegraphics[width=11cm]{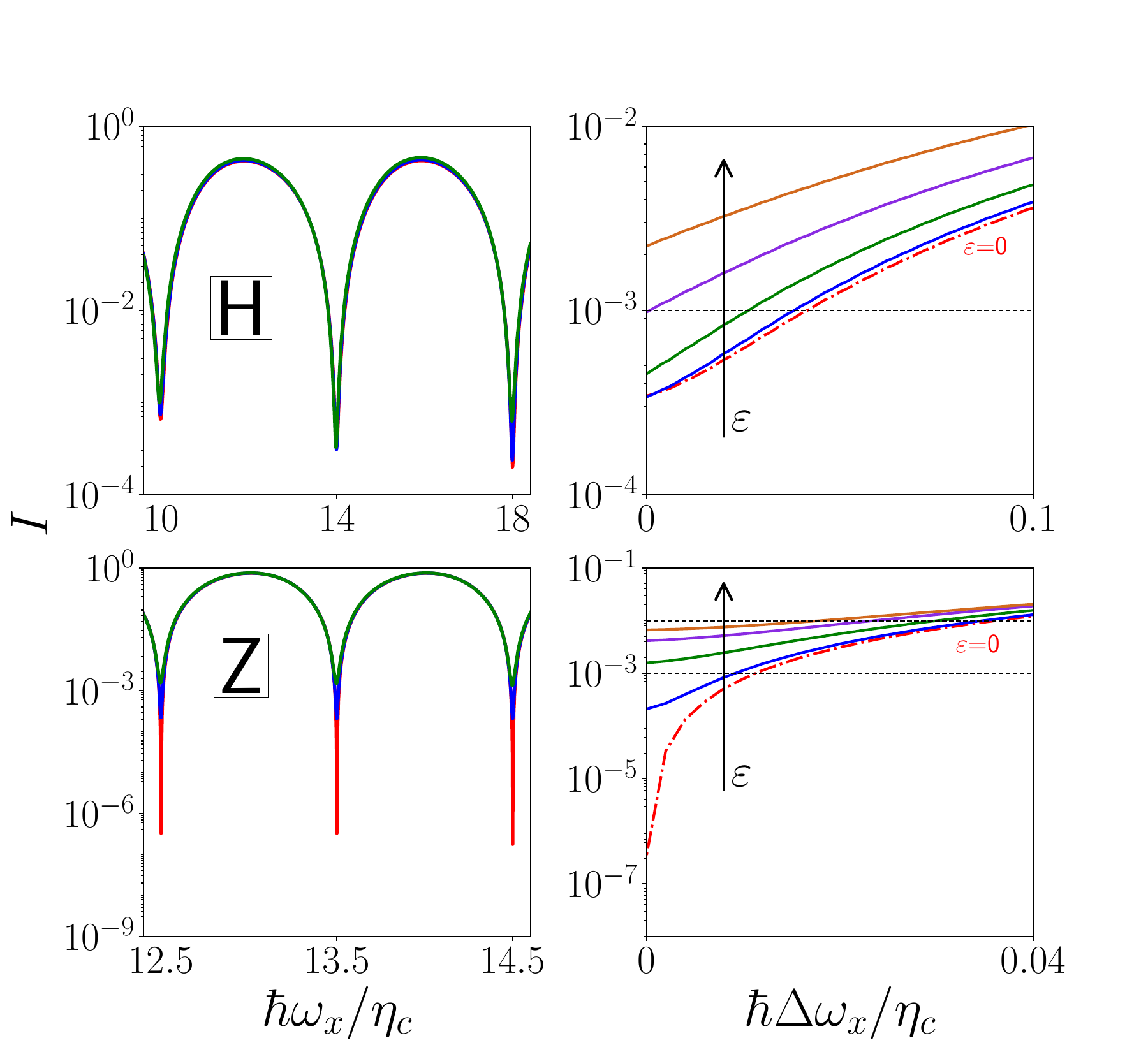}
    \caption{\label{fig:H-Z-gates} Left: Infidelity $I$ for $H$ and $Z$ gates as a function of $\hbar\omega_x/\eta_c$ for level detuning $\epsilon=0$, 0.2 and 0.4 $\mu$eV. Right: Infidelity for $H$ and $Z$ gates as a function of $\hbar\Delta\omega_x/\eta_c$ for values of detuning $\epsilon$ increasing from zero (lower curve) to 0.8 $\mu$eV (higher curve) through steps of 0.2 $\mu$eV as indicated by the upward arrow. $\Delta \hbar\omega_x/\eta_c$ are shifts from the central dip $\hbar\omega_x/\eta_c$ on the left panel ($B_x=150$ mT, and $\eta_c=0.643\ \mu$eV).}
\end{figure}

Fig.  \ref{fig:H-Z-gates}a shows the calculated infidelity $I$ of the numerically computed unitary evolution $U_{\rm num}$ with respect to the ideal analytical gates, at detuning $\epsilon=0$, 0.2 and 0.4 $\mu$eV, as a function of the parameter $\hbar\omega_x/\eta_c$ for $H$ and $Z$ gates.
Strong dips of low $I\lesssim 10^{-3}$ occur at the values expected from the analytical model ($n_{\rm odd}=5$, 7, 9 for $H$, and $n_{\rm odd}=25$, 27, 29 for $Z$ are shown). In between those values $I$ becomes large, in the range 0.1--1, where the evolution would departure from the ideal gates. 

Fig. \ref{fig:H-Z-gates}b shows the infidelity $I$ as a function of the shift $\hbar\Delta\omega_x/\eta_c$ from the central dip. This quantity is a measure of the off-resonant shift $\Delta\omega_x$ with respect to the Rabi frequency $\eta_c/\hbar$. 
Different curves correspond to various magnitude of detuning from zero to 1 $\mu$eV increasing upwards, as indicated by the arrow. Most of the curves are in the range $I\lesssim 10^{-2}$ and a horizontal dashed line marks the usual threshold $I=10^{-3}$.

\begin{figure}[H]
    \centering
    \includegraphics[scale=0.3]{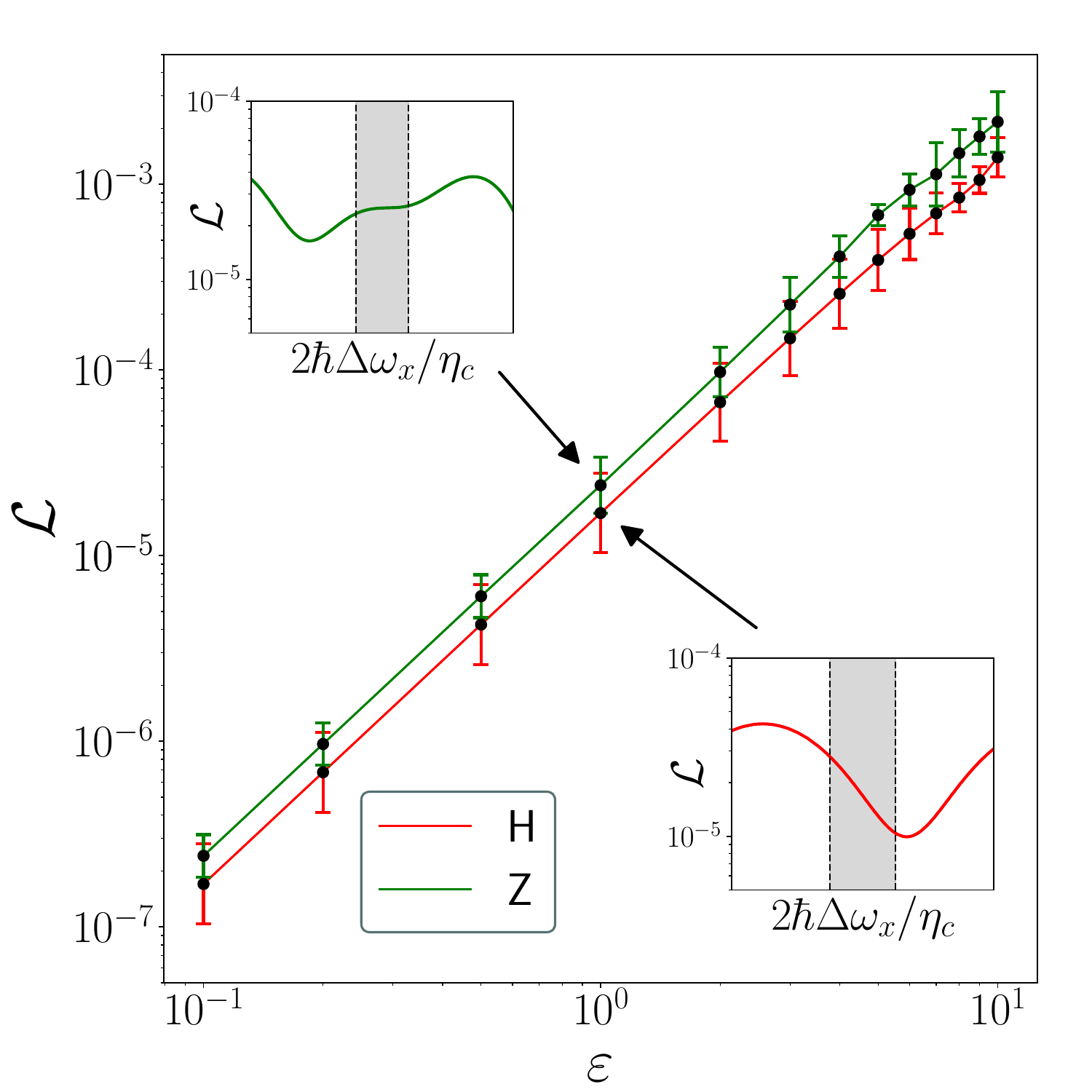}
    \caption{\label{fig:leakage} Numerically calculated leakage from the dynamics of the exact Hamiltonian corresponding to Hadamard $H$ and $Z$ gates as a function of the detuning $\epsilon$ with parameters set from the analytically defined gates. The longitudinal and transverse magnetic fields are $B_z=600$ mT and $B_x=150$ mT, respectively. Insets depict the leakage as a function of shifts from resonance (in units of $\eta_c$) for the two gates at $\epsilon=1\ \mu$eV as pointed by the arrows. The gray shaded areas correspond to the same ranges shown in Fig. \ref{fig:H-Z-gates}.}
\end{figure}

Fig. \ref{fig:leakage} shows the leakage of the unitary evolutions representing $H$ and $Z$ gates, as a function of detuning for a wide range of values, $0.1\leq \epsilon\leq 10\ \mu$eV. The symbols are the calculated mean values of ${\cal L}$ during the time of operation of the gate, the error bars shows its range of variation during the evolution and the solid line joining the symbols is only drawn as a guide. The approximated linear log-log plot indicates a power-law relation ${\cal L}\sim \epsilon^2$ for both gates. 
The insets show ${\cal L}$, at a typical detuning $\epsilon=1\ \mu$eV for each gate, as a function of the parameter of shift from resonance $\hbar\Delta\omega_x/\eta_c$. The vertical gray fringes correspond to the same range of values used in Fig. \ref{fig:H-Z-gates}.
\section{Two-qubit gates \label{2-qubit gates}}
\subsection{Unitary evolution by exchange coupling modulation}
Now consider the dynamics of the effective two-electron Hamiltonian, Eq. (\ref{H6x6}), with a time-dependent exchange coupling $J(t)$. We shall show that the time evolution operator $U_2(t)$ will generate the required two-qubit operations under proper parameter choices. 


The dynamics of a general state $i\partial_t\psi=H\psi$ evolves along two separate 3-dimensional subspaces, namely, ${\cal H}_\Phi={\rm span} \{|00\rangle, |11\rangle, |\chi_1\rangle\}$ and ${\cal H}_\Psi= {\rm span}\{|10\rangle, |01\rangle, |\chi_2\rangle\}$. Within these subspaces each pair of computational states is coupled to one another and to a leakage state $|\chi_{1}\rangle$ or $|\chi_{2}\rangle$.

By expanding the two-electron state as 
\begin{equation}
    |\psi\rangle = \sum_{k\in S} a_k(t) \exp(-i\omega_k t/2) |k\rangle,
\end{equation}
where $S=\{00,10,01,11,\chi_1,\chi_2\}$ is the set of labels of states and the frequencies are $\omega_{\chi_1}=\omega_{\chi_2}=-J$, $\omega_{00}=-\omega_{11}=\Omega_1$ and $\omega_{10}=-\omega_{01}=\Omega_2$, the respective Schr\"odinger equations takes the form

\begin{equation} \hspace*{-2.5cm}
    i\left(
    \begin{array}{c}
         \dot{a}_0 \\
         \dot{a}_1 \\
         \dot{b}
    \end{array}
    \right) = \frac{J(t)}{4}\left(
    \begin{array}{ccc}
        0 & e^{i\omega t} & \sqrt{2} e^{i(J/2+\omega) t/2}  \\
         e^{-i\omega t} & 0 & - \sqrt{2} e^{i(J/2-\omega) t/2}  \\
        \sqrt{2} e^{-i(J/2+\omega) t/2} & - \sqrt{2} e^{-i(J/2-\omega) t/2} & 0
    \end{array}
    \right)
    \left( 
    \begin{array}{c}
         a_0 \\
         a_1 \\
         b
    \end{array}
    \right)
    \label{eq:3D-dynamics}
\end{equation}
where the probability amplitudes and frequencies are $(a_0, a_1, b, \omega) =(a_{00}, a_{11}, a_{\chi_1}, \Omega_1)$ for ${\cal H}_\Phi$, while $(a_0, a_1, b, \omega) =(a_{10}, a_{01}, a_{\chi_2}, \Omega_2)$ for ${\cal H}_\Psi$.\\

We propose to modulate harmonically the exchange coupling with two frequencies, $\omega_1$ and $\omega_2$, to operate within each subspace, as 
\begin{equation}
    J(t) = J_0 + J_1 \sin \omega_1 t+J_2 \sin \omega_2 t,
    \label{eq:J-modulation}
\end{equation}
where $J_0$ is a static exchange coupling, and $J_1$ and $J_2$ are the amplitudes of the biharmonic modulation superposed such that $J_1+J_2\leq J_0$.

Eq. (\ref{eq:3D-dynamics}) can be solved within RWA by comparing the frequencies of the non-interacting electrons, $\Omega_i$, with the excitation frequencies of oscillating exchange, $\omega_j$; in this approximation, rapidly oscillating terms having frequencies $\Omega_i+\omega_j$ and $\Omega_i/2+\omega_j$ are neglected.

Terms with difference of frequencies $\delta=\Omega_i-\omega_j\simeq 0$ are the main contribution and are conserved. Terms with frequencies $\Omega_i/2-\omega_j=\delta-\Omega_i/2$ give small corrections. The parameter $\delta$ is the frequency detuning of the modulation with respect to the natural frequencies $\Omega_1$ and $\Omega_2$ in ${\cal H}_\Phi$ and ${\cal H}_\Psi$, respectively.\\

With these approximations the equations become
\begin{equation} \hspace*{-0.5cm}
 i\left(
    \begin{array}{c}
         \dot{a}_0 \\
         \dot{a}_1 \\
         \dot{b}
    \end{array}
    \right) = \frac{i}{8}\left(
    \begin{array}{ccc}
       0 &  J_ie^{i\delta t} &  \sqrt{2}J_je^{-i(\omega_j-\Omega_i^+) t}   \\
        -J_ie^{-i\delta t} & 0 & -\sqrt{2}J_je^{i(\omega_j-\Omega_i^-) t} \\
       -\sqrt{2}J_je^{i(\omega_j-\Omega_i^+) t} &  \sqrt{2}J_je^{-i(\omega_j-\Omega_i^-) t} & 0
    \end{array}
    \right) \left( 
    \begin{array}{c}
         a_0 \\
         a_1 \\
         b
    \end{array}
    \right),
    \label{eq:3-dim-dynamics}
\end{equation}
where $\Omega_i^{\pm}=\Omega_i/2\pm J/4$, such that at resonance ($\delta=0$) and neglecting terms of frequency $\pm\Omega_i/2$, the leakage is suppressed ($\dot{b}=0$) and the probability amplitudes $a_0(t)$ and $a_1(t)$ oscillate harmonically with frequency $J_i/8$. 
\begin{equation}
    \left(
    \begin{array}{c}
        a_{0}e^{-i\Omega_i t/2}  \\
        a_{1}e^{i\Omega_i t/2}
    \end{array} 
    \right) =
    \left(
\begin{array}{cc}
    \cos (J_i t/8) & \sin (J_i t/8) \\
     -\sin (J_i t/8) & \cos (J_i t/8)
\end{array}
    \right) \left(
    \begin{array}{c}
        a_{0}(0)  \\
        a_{1}(0)
    \end{array} 
    \right),
\end{equation}

Then, the evolution operator in the computational basis $\{|00\rangle, |10\rangle, |01\rangle, |11\rangle\}$ becomes 
\begin{equation}
    \hspace*{1.8cm} 
{\renewcommand{\arraystretch}{1.2} 
\resizebox{0.78\textwidth}{!}{$
\fl    U_2 =  \left(
\begin{array}{cccc}
        \cos (J_1t/8) e^{-i\Omega_1 t/2}& 0 & 0 & \sin (J_1t/8) e^{-i\Omega_1 t/2} \\
               0 &     \cos (J_2t/8) e^{-i\Omega_2 t/2}& \sin (J_2t/8) e^{-i\Omega_2 t/2}  & 0 \\
               0 &    -\sin (J_2t/8) e^{i\Omega_2 t/2} & \cos (J_2t/8) e^{i\Omega_2 t/2}    & 0 \\
         -\sin (J_1t/8) e^{i\Omega_1 t/2} & 0 & 0 & \cos (J_1t/8) e^{i\Omega_1 t/2} 
\end{array}
\right).
\label{eq:2-qubit-U}
$}}
\end{equation}

The unitary evolution operator, Eq. (\ref{eq:2-qubit-U}), can provide entangling gates by suitably choosing the amplitudes $J_1$ and $J_2$. 
Different two-qubit gates are equivalent to each other, up to local operations, if they have the same Makhlin invariants $G_1$ and $G_2$  \cite{Makhlin02}:
\begin{eqnarray}
    G_1 &=& \frac{1}{16} {\rm Tr}^2 m(t) \det U^\dag (t) \\
    G_2 &=& \frac{1}{4} \left[{\rm Tr}^2 m(t) - {\rm Tr\ } m^2(t)\right] \det U^\dag (t),
\end{eqnarray}
where $m(t) = M^T M $ and $M(t)$ is the transformation of the evaluated unitary transformation to Bell basis
\begin{equation}
    M(t) = Q^\dag U(t) Q,
\end{equation}
with
\begin{equation}
    Q = \left(
    \begin{array}{cccc}
        1 & 0 & 0 & i \\
        0 & i & 1 & 0 \\
        0 & i & -1 & 0 \\
        1 & 0 & 0 & -i 
    \end{array}
    \right).
\end{equation}
Makhlin invariants calculated for the two-qubit evolution operator, Eq. (\ref{eq:2-qubit-U}), are
\begin{eqnarray}
    G_1(t) &=& \frac{1}{4}\left[\cos (J_1t/4)+\cos(J_2t/4)\right]^2\\
    G_2(t) &=& 1+2 \cos (J_1t/4) \cos(J_2t/4).
\end{eqnarray}
Note that they become independent on the frequencies $\Omega_1$ and $\Omega_2$. By properly choosing the operation time and the $J_2/J_1$ relation various two-qubit gates are obtained. The values of $ G_1$, $G_2$ and the locally equivalent gates for $U_2(J_1, \Omega_1,J_2, \Omega_2,t) $ are given in Table \ref{tab:2-qubit gates}. Next, we shall concentrate on its operation as a CNOT-equivalent gate.
\begin{table}[]
    \centering
    \begin{tabular}{ccccc}
    \hline\hline
        $J_2/J_1$ & $T$ & $G_1$ & $G_2$ & Gate \\ \hline
              1   & $2\pi/J_1$ & 0 & 1  & {\rm CNOT} \\
              2   & $4\pi/J_1$ & 0 & -1 & $i${\rm SWAP} \\
              2   & $2\pi/J_1$ & 1/4 & 1 & $\sqrt{i{\rm SWAP}}$ \\ \hline \hline
    \end{tabular}
    \caption{Relation $J_2/J_1$ of the amplitudes of modulation, time of operation $T$ and Makhlin invariants $G_1$ and $G_2$ for realizable two-qubit gates with the evolution operator $U_2(J_1, \omega_1,J_2, \omega_1,t)$, Eq. (\ref{eq:2-qubit-U}). }
    \label{tab:2-qubit gates}
\end{table}
\subsection{Numerical calculations. \label{two-qubit simulation}}

The proposed analytical approximations for realizing CNOT-equivalent entangling gates set conditions on the unitary evolution: (i) resonant biharmonic modulation of exchange interaction at the frequencies $\Omega_1$ and $\Omega_2$ of the non-interacting electron levels, (ii) equal amplitudes for $J_1=J_2$ for both frequencies, (iii) negligible coupling $\Delta\Omega$ between leakage states. We assessed the sensitivity of the dynamics to those conditions through numerical simulations using the exact Hamiltonian.

\begin{figure*}
\centering
$
\begin{array}{c}
\includegraphics[width=9.5cm]{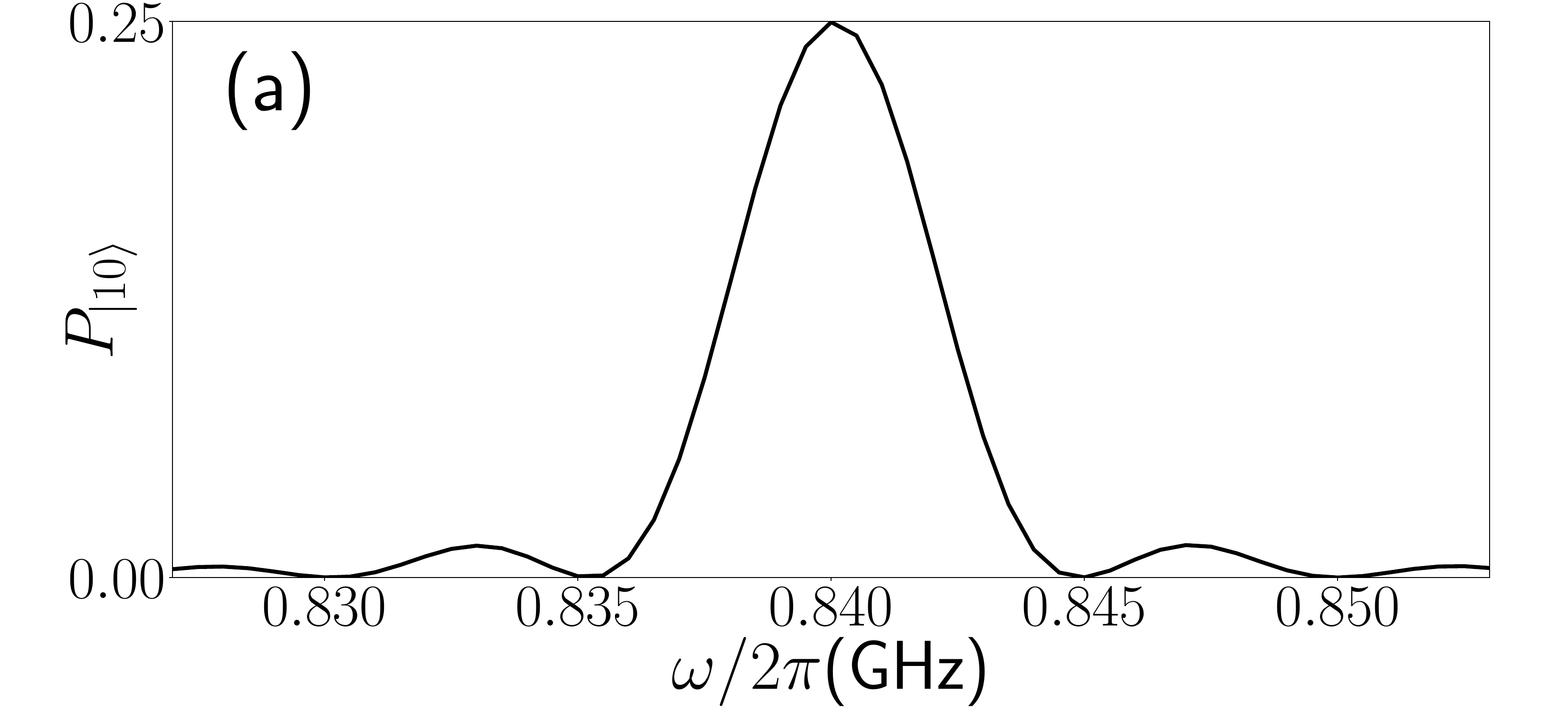} \\
\begin{array}{cccc}
\includegraphics[width=6cm]{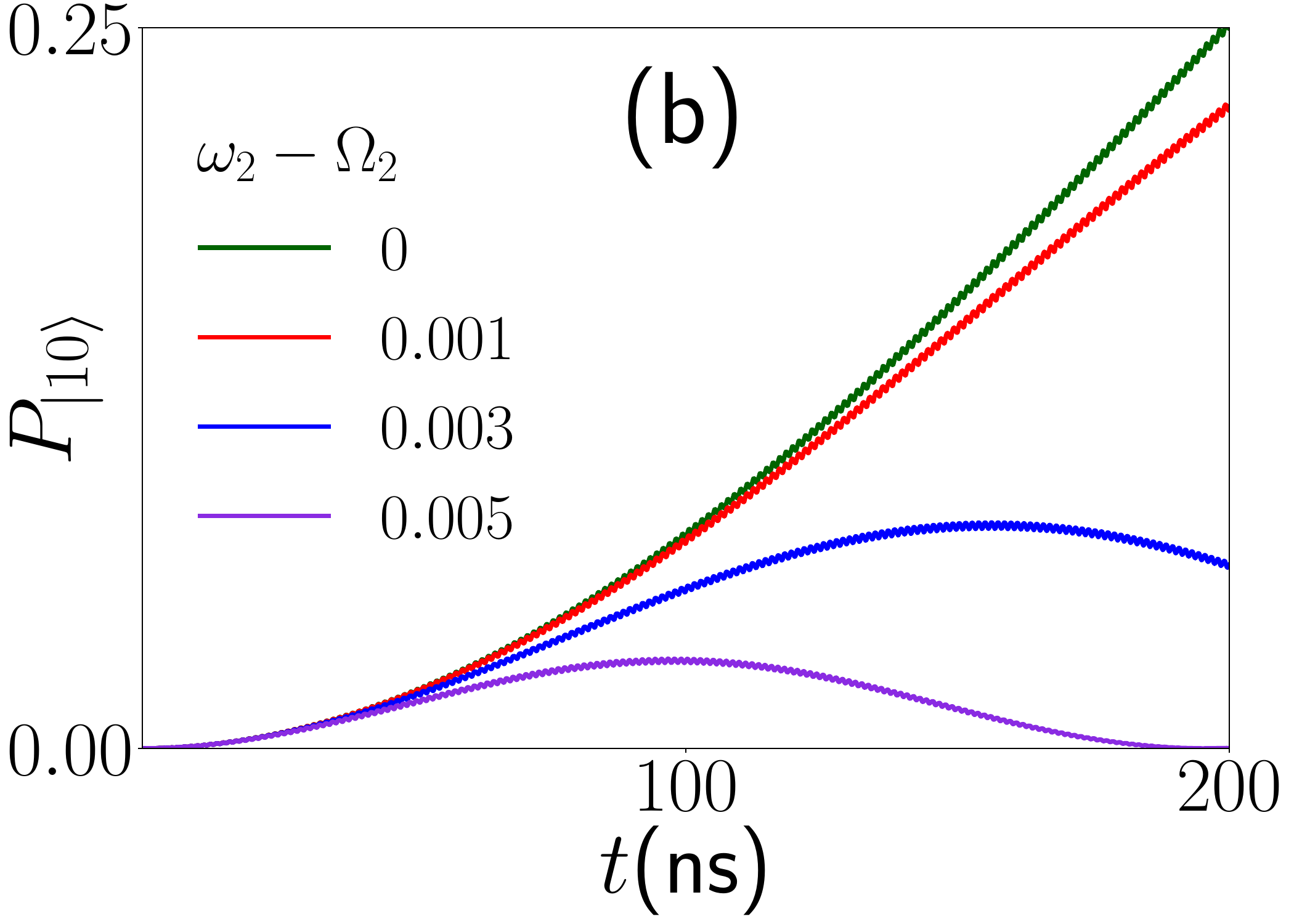} &
\includegraphics[width=6cm]{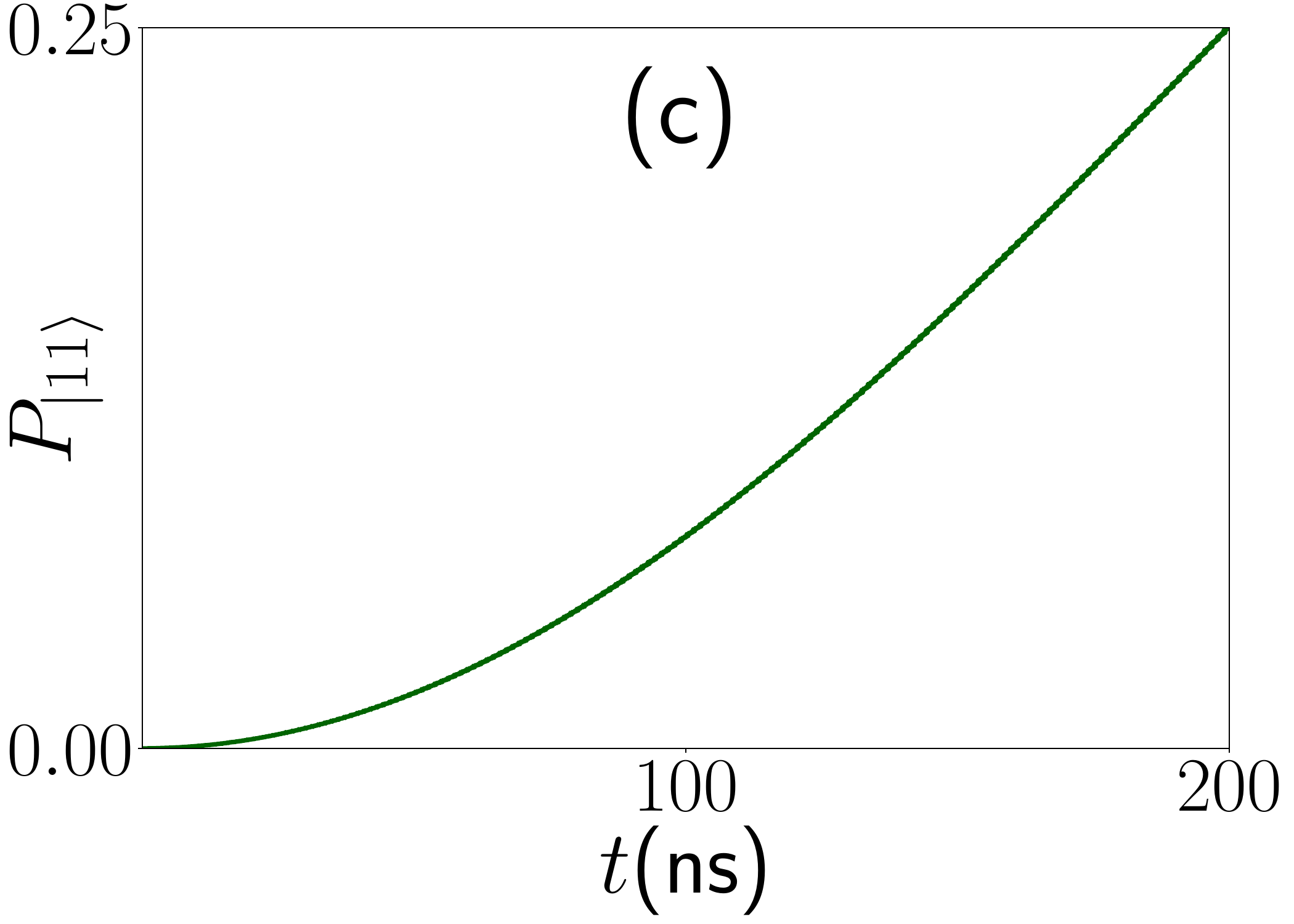} &
\end{array}
\end{array}$
 \caption{\label{fig:dynamics-2-qubit} (a) Occupation of state $|10\rangle$ calculated as a function of frequency of modulation $\omega_2$, after unitary evolution from $|\psi(0)\rangle=(|00\rangle+|01\rangle)/\sqrt{2}$, with biharmonic coupling $J(t)=J_0+J_1 (\sin\omega_1 t+ \sin\omega_2 t)$, Eq. (\ref{eq:J-modulation}) with $J_1=J_2$ for a CNOT-equivalent entangling gate, for transverse fields $(B_x^{(1)},B_x^{(2)})=$ (120, 60) mT, giving resonant frequencies $\Omega_1/2\pi=2.52$ GHz and $\Omega_2/2\pi=0.84$ GHz; amplitudes of control are $J_0=2\pi\times 10$ MHz and $J_1=J_2=2\pi\times 5$ MHz. (b)-(c) Occupation of state $|10\rangle$ y $|11\rangle$ as a function of the evolution time $t$ in nanoseconds. The frequency $\omega_1$ is resonant at $\Omega_1/2\pi=2.52$ GHz, while $\omega_2$ if off-resonant: $\omega_2/2\pi=0.84$, 0.841, 0.843 and 0.845 GHz. Initial state $|\phi_+\rangle=\frac{1}{\sqrt{2}}(|00\rangle+|01\rangle)$.} 
\end{figure*}

Firstly, we present numerical results for the exact time evolution of a state driven by the exchange interaction modulated biharmonically. The calculations were performed with $B_x^{(1)}=$ 120 mT and $B_x^{(2)}=$ 60 mT, i.e.  frequencies $\Omega_1/2\pi=2.52$ GHz and $\Omega_2/2\pi=0.84$ GHz; the control amplitudes were set to $J_0=2\pi\times 10$ MHz and $J_1=J_2=2\pi\times 5$ MHz. 
Since the resonant conditions within both subspaces are needed for the qubit evolution to match a CNOT-equivalent gate, we assessed the effect of setting on of them ($\omega_2$) slightly off-resonant. We initialize the two-qubit state as a superposition of states in ${\cal H}_\Phi$ and ${\cal H}_\Psi$, $|\psi(0)\rangle=(|00\rangle+|01\rangle)/\sqrt{2}$, and numerically calculate its evolution along the ideal gate operation time $T_{\rm op}=2\pi/J_1=200$ ns. At that time, the population should be equally distributed among the four two-qubit computational states. Fig. \ref{fig:dynamics-2-qubit}(a) shows the population $P_{|10\rangle}=|\langle 10|\psi(T)\rangle|^2$ as a function of $\omega_2$ in a range around the resonant frequency $\Omega_2$, while $\omega_1=\Omega_1$ is kept resonant. The occupation of $|10\rangle$ in ${\cal H}_\Psi$, controlled by $\omega_2$, reaches $1/4$ at resonance but is sensitive to a frequency detuning of a few Hz, Fig. \ref{fig:dynamics-2-qubit}(b). The states in ${\cal H}_\Phi$, excited resonantly at $\omega_1=\Omega_1$, nevertheless, perform the derived Rabi oscillations with frequency $J_1/8$, Fig. \ref{fig:dynamics-2-qubit}(c).  Then, the approximate analytical dynamics of the two-qubit state, invoked to derive the entangling gate, is closely satisfied by the exact evolution, and it can be controlled by independently tuning the frequencies and amplitudes in $J(t)$.
Now we turn to the analysis of the infidelity and leakage in the two-qubit gates. The unitary evolution can fail to generate the designed entangling gate due to the frequency detuning induced by the exchange interaction. It contains the static contribution $J_0$, and rapidly varying time-dependent terms that can be neglected assuming they average to zero. Then, the qubit energy levels shift from $\Omega_i$ to $\widetilde{\Omega}_{i}=\sqrt{\Omega_i^2+ J_0^2/4}\approx \Omega_i +  J_0^2/8\Omega_i$. This also introduces some hybridization in the qubit states defined from the independent DQD energy levels that we disregard. The extra term induce a phase shifts in the elements of the unitary evolution $U_2$, Eq. (\ref{eq:2-qubit-U}), through $\exp(\pm i\widetilde{\Omega}_iT_{\rm op}/2)\approx \exp(\pm i\Omega_iT_{\rm op}/2)\exp(\pm i\pi  J_0^2/4J_1)$. Therefore, the infidelity $I$ can be expected to approximately depend as


\begin{equation}
I(J_0,J_1)=1-|\langle\psi(0)|U^\dagger(\Omega_i) U(\widetilde{\Omega}_i)|\psi(0)\rangle|^2 \approx a  \left( \frac{J_0^2}{J_1} \right)^2,     
\label{I-analytical}
\end{equation} 
with $a$ independent on $J_0$ and $J_1$. Furthermore, a similar argument sets the minimal value of $I$ for $J_0=2J_1 $, giving
\begin{equation}
I_{min}(J_1)\simeq   4 a  J_1^2 = 16 \pi^2 a \left( \frac{1}{T_{\rm op}}\right)^2,  
\end{equation} 
which shows that reducing the operation time contributes quadratically to improve the gate fidelity.

\begin{figure*}
    \centering
    $\begin{array}{c}
     \includegraphics[width=14cm]{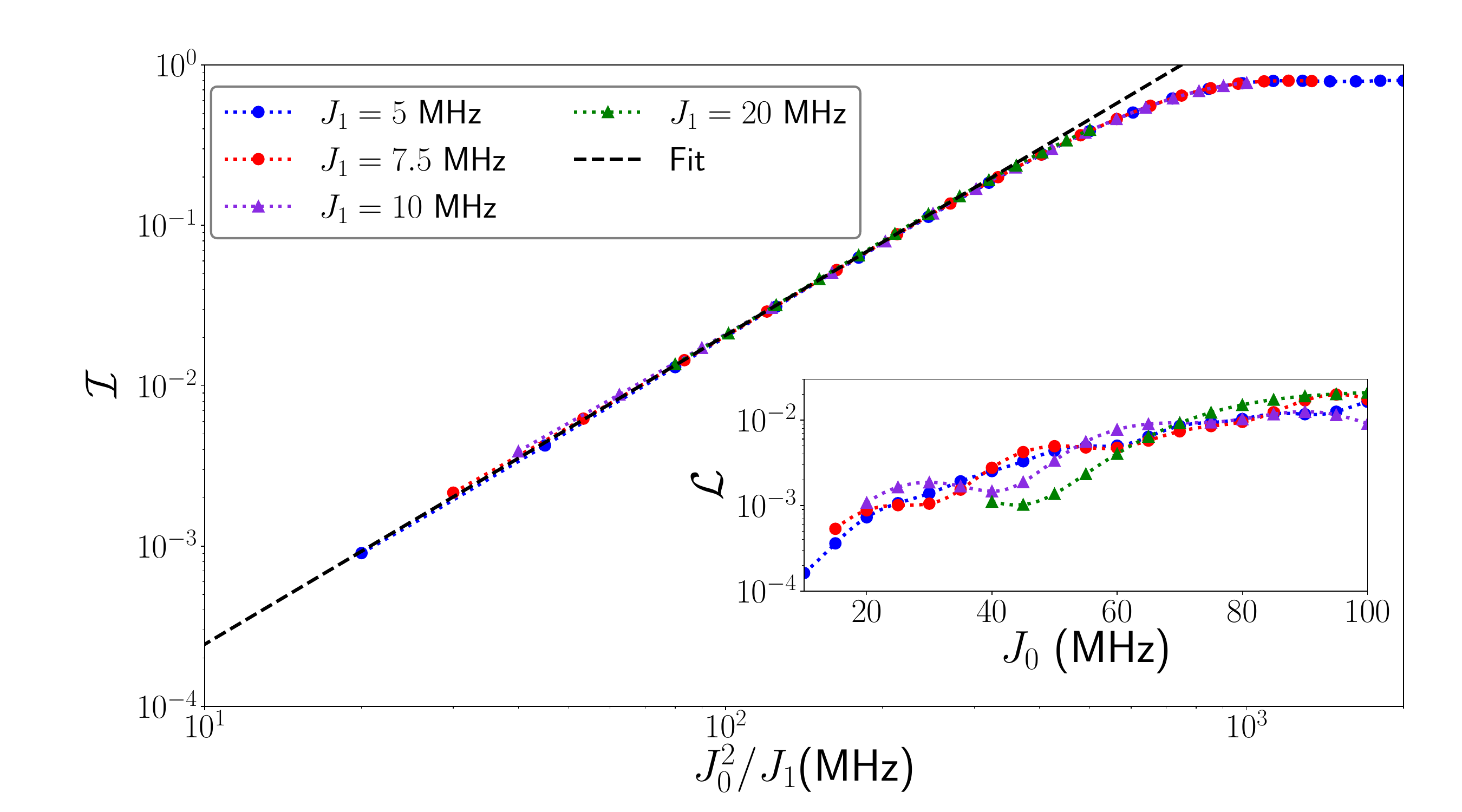} \\
     \begin{array}{cc}
          \includegraphics[width=7cm]{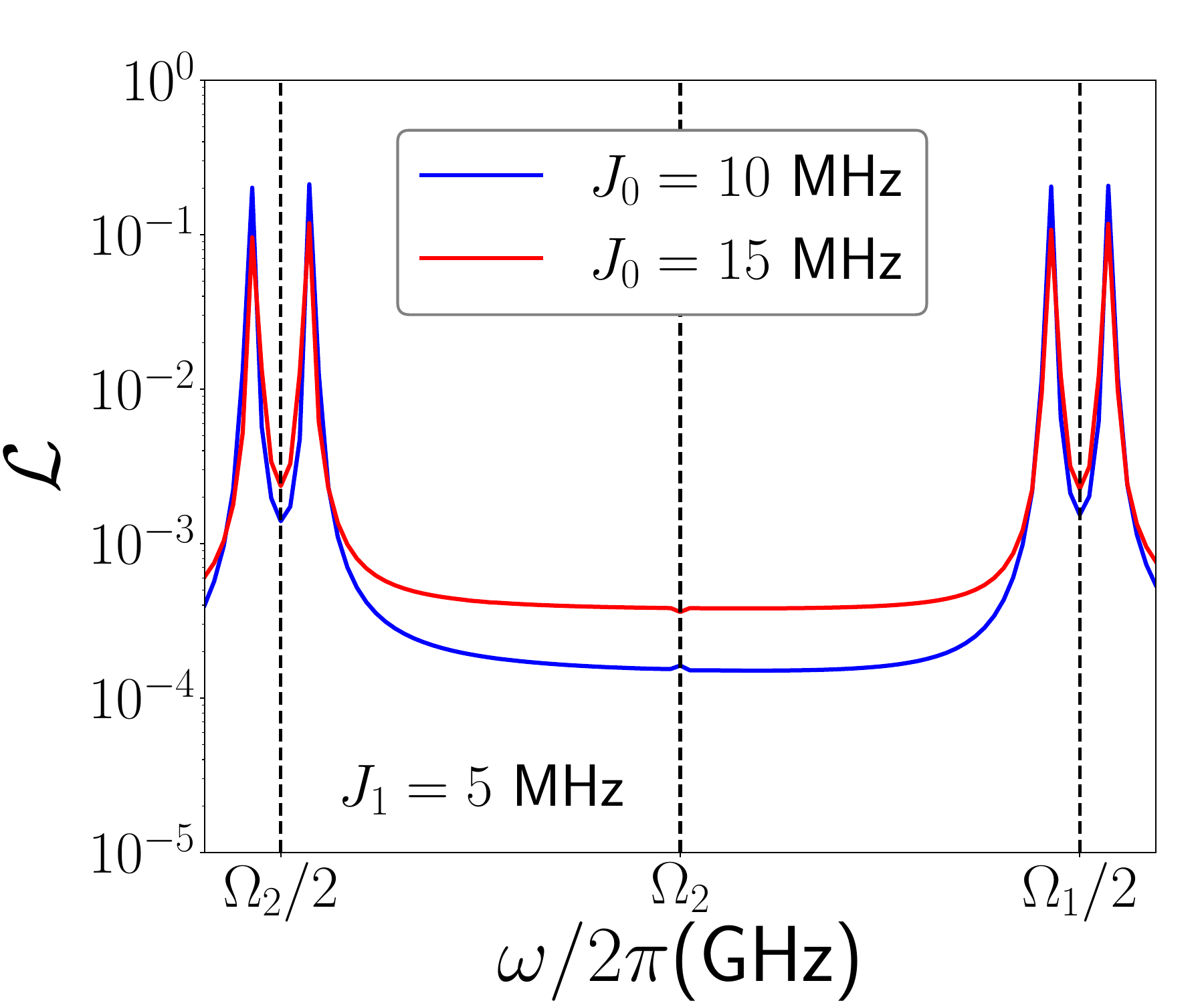} &
          \includegraphics[width=7cm]{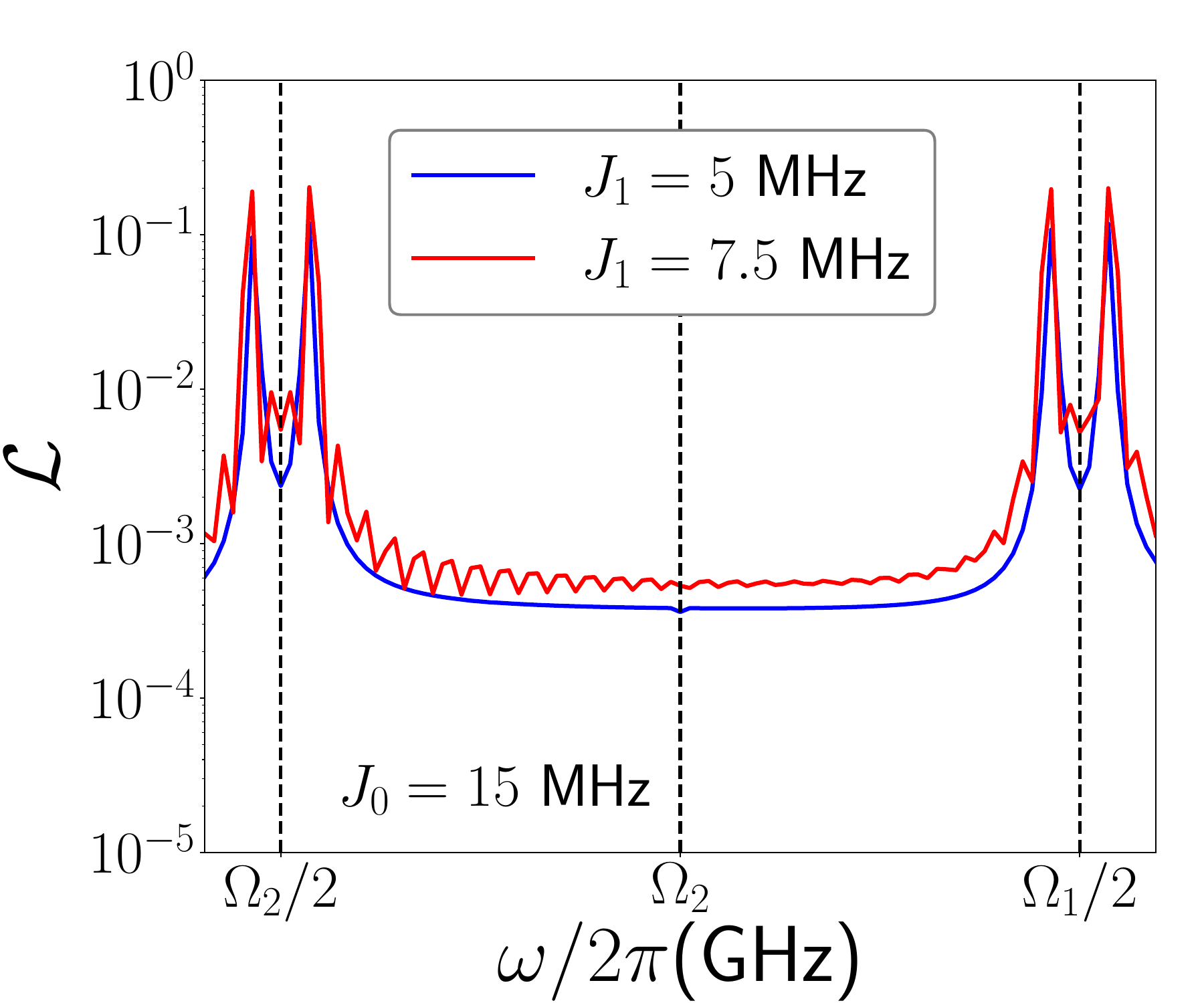}
     \end{array}\\
    \end{array}$
    \caption{
    \label{fig:I-leakage 2-qubit} Calculated infidelity $I$ and leakage ${\cal L}$ for the two-qubit entangling gate. (\textit{Above}) $I$ as function of the $J_0^2/J_1$. The dashed straight line is the fitting $a x^n$, where $x=J_0^2/J_1$ and the parameters are: $a=3\times 10^{-6}$ MHz$^{-2}$, $n=2.0$. The \textit{Inset} shows ${\cal L}$ as function of the $J_0$. (\textit{Below}) ${\cal L}$ as function of the $\omega_2$ for the different values of the $J_0$ and $J_1$. The dashed line in the middle of the graph corresponds to the resonance point: $\omega_2=\Omega_2$.}
\end{figure*}

In order to test the analytical approximations, we performed numerical calculations with the same set of parameters given above. We show in Fig. \ref{fig:I-leakage 2-qubit}(a), the results of calculated infidelity $I$, using Eq. (\ref{eq:def_Fidelity}), for numerically calculated evolutions versus $J^2_0/J_1$, for several values of $J_1$. The linear range of the plot shows a power-law relation that is well fitted by $I\simeq 3\times 10^{-6} {\rm \ MHz}^{-2} (J_0^2/J_1)^2$, in agreement with the analytical estimation, Eq. (\ref{I-analytical}). The departures from power-law dependence correspond to values of $J_0$ high enough to deteriorate the behavior of the system as a two-qubit gate. The inset shows the corresponding calculated leakage ${\cal L}$, using Eq. (\ref{eq:def_Fidelity}), and its approximate ${\cal L}\sim J_0^2$ dependence.



Similarly, we studied the effect of frequency detuning on leakage. We assumed $J(t)$ resonant in subspace ${\cal H}_\Phi$, $\omega_1=\Omega_1$, and numerically calculated ${\cal L}$ as a function of $\omega_2$ varying on a wide range around the ${\cal H}_\Psi$ resonance ($\omega_2=\Omega_2$). Unwanted transitions to leakage states are analytically expected to occur at frequencies, Eq. (\ref{eq:hamiltonian-2qb}), 
\begin{equation}
    \omega_{\rm leak} = \pm\left( \frac{\Omega_i}{2} + \frac{J_0}{4}\right) \pm \Delta\Omega.
\end{equation}
Figs. \ref{fig:I-leakage 2-qubit} (b) and \ref{fig:I-leakage 2-qubit} (c) show the numerical results. Near the resonance $\omega_2=\Omega_2$, the leakage is flat and small ($\lesssim 10^{-3}$), highly advisable for the correct working of the device. Peaks of high leakage is observed around $\omega_2=\pm\Omega_2/2$ split by $\Delta\Omega\ll\Omega_2$, in agreement with the analytical result. This region of poor performance of the device is far away from the proposed working range. Hence, no significant deterioration is expected due to such effect.
Notably, none of those peaks lie close to the operation point, as long as the two applied frequencies ratio is not near to one-half or two. Since that result requires a very specific relation between $\Omega_1$ and $\Omega_2$, the validity of the RWA approximation is ensured within a small margin of infidelity defined by $J_0$ and $J_1$, as discussed above.
\section{Conclusions \label{conclusions}}
In this work, we proposed to use as qubits the states of a system of two electrons in two electrostatically defined double quantum dots in a semiconductor heterostructure, subject to the magnetic field of a micromagnet. The working point is set at zero level detuning (symmetrical quantum dot ground states) to decouple the qubits from leakage levels, and at a static tunneling gap to exactly compensate the longitudinal Zeeman splitting and isolate the levels of interest. 

We have shown analytically that the complete set of one- and two-qubit operations can be produced by the harmonic modulation of the tunnel coupling in the double quantum dots, and the exchange interaction between them.
One-qubit gates are generated by resonant modulation of the tunneling  around the working point at the frequency $\omega_x$ of the Zeeman gap produced by the micromagnet transverse magnetic field at a single DQD. A longitudinal gradient of $B_x$ allows to distinguish resonant frequencies at each DQD. 
Two-qubits gates are produced by the harmonic modulation of the exchange coupling $J(t)$ with frequencies resonant at the sum ($\Omega_1$) and the difference ($\Omega_2$) of the one-qubit Zeeman frequency gap of each DQD. 
Then, the two-qubit subspaces ${\cal H}_\Phi$ or ${\cal H}_\Psi$ are independently controlled by $\Omega_1$ and $\Omega_2$, allowing to generate entangled states.

We also have performed numerical calculations with the exact Hamiltonian, to assess the sensitivity of the analytically defined operations to imperfect matching to the ideal parameters. 
The results show that the small leakage at resonant frequencies, and the controllability predicted by the analytical model, are still fulfilled within reasonable shifts from the sweet spots. 
The difference in the energy scales of tunneling ($\mu$eV) and exchange coupling ($\sim 10^{-2}\ \mu$eV) results in the fact that one-qubit gates are much faster ($\sim$ ns) than two-qubit ones ($\sim 10^2\ $ns). 
The proposed qubit provides a different way to use coupled double quantum dots for quantum computing operations.

\section*{acknowledgments}
Authors acknowledge financial support from grants PUE2017-22920170100089CO (CONICET, Argentina), PI-20T001 and PI-23F005 (Universidad Nacional del Nordeste, Argentina).
\appendix

\section{Derivation of the 6-level Hamiltonian, Eq. (10)}
We use the two-spin basis $\{|S_{ij}\rangle,|T_{ij}^0\rangle,|T_{ij}^+\rangle,|T_{ij}^-\rangle\}$ within each block, where $|S_{ij}\rangle$ and $|T_{ij}^0\rangle$ are the singlet (${\cal S}=0$)  and triplet (${\cal S}=1$) unpolarized (${\cal S}_z=0$) states $(|\uparrow_i\downarrow_j\rangle\pm |\downarrow_j\uparrow_i\rangle)/\sqrt{2}$, while $|T_{ij}^+\rangle=|\uparrow_i\uparrow_j\rangle$ and $|T_{ij}^-\rangle=|\downarrow_i\downarrow_j\rangle$ are the corresponding polarized (${\cal S}_z=\pm 1$) triplet configurations, between bonding and antibonding states $i,j\in \{a,b\}$ from each qubit. This basis is orthonormal both in the spacial and spin degrees of freedom, i.e., $\langle {\cal S}_{ij}^{m}|{\cal S'}_{kl}^{m'}\rangle= \delta_{{\cal S}{\cal S}'}\delta_{m m'} \delta_{ik} \delta_{jl}$, where ${\cal S}$, ${\cal S'}$ are the total spin momentum, $m$, $m'$ are their $z$-component, and $i, j, k, l \in \{a,b\}$. Assuming a vanishing overlap of spacial wave functions localized at different QDs, any pair of orbitals (either bonding and antibonding) are orthogonal if they belong to different qubits, and $\langle a|b\rangle=0$ within each qubit.

In this spin-adapted basis the operator that exchanges spins ($P_{12}|\sigma_1\sigma_2\rangle=|\sigma_2\sigma_1\rangle$) takes the form $P_{12}=(\bm{\sigma}_1 \cdot \bm{\sigma}_2+1)/2$  and satisfy

\begin{equation}
\begin{split}
    \bm{\sigma}_1 \cdot \bm{\sigma}_2 |S_{ij}\rangle &= -|S_{ij}\rangle-2|S_{ji}\rangle,\\
    \bm{\sigma}_1 \cdot \bm{\sigma}_2 |T^m_{ij}\rangle &= -|T^m_{ij}\rangle+2|T^m_{ji}\rangle, \quad (m=0,\pm 1)
     \label{exchange-property}
\end{split}
\end{equation}
when applied to an arbitrary singlet or triplet state. Note the swapping in spatial orbital indices $i$ and $j$ such that, if both qubits are in the same orbital $i$, $\bm{\sigma}_1\cdot\bm{\sigma}_2|S_{ii}\rangle=-3 |S_{ii}\rangle $ and $\bm{\sigma}_1\cdot\bm{\sigma}_2|T^m_{ii}\rangle= |T^m_{ii}\rangle$, i.e,  they are eigenstates of $H_{\rm exch}$.

Then, the two-qubit logical space ${\cal H}_{\rm comp}^{\otimes 2}$ is spanned by the orthonormal basis 
\begin{align}
    |00\rangle &= \frac{1}{2} |T_{aa}^+\rangle + \frac{1}{2} |T_{bb}^-\rangle + \frac{1}{\sqrt{2}} |T_{ab}^0\rangle \\
    |11\rangle &= \frac{1}{2} |T_{aa}^+\rangle + \frac{1}{2} |T_{bb}^-\rangle - \frac{1}{\sqrt{2}} |T_{ab}^0\rangle \\
    |01\rangle &= \frac{1}{2} |T_{aa}^+\rangle - \frac{1}{2} |T_{bb}^-\rangle - \frac{1}{\sqrt{2}} |S_{ab}\rangle \\
    |10\rangle &= \frac{1}{2} |T_{aa}^+\rangle - \frac{1}{2} |T_{bb}^-\rangle + \frac{1}{\sqrt{2}} |S_{ab}\rangle,
    \label{2-qubit-basis}
\end{align}

The leakage subspace ${\cal H}_{\rm leak}^{\otimes 2}$ can also be given as combinations of spin-adapted states.


Finally, transforming to the basis $\{|\chi_k\rangle\}$:  
\begin{equation}
\begin{split}
    |\chi_1\rangle = \frac{1}{\sqrt{2}}\left(|l_1l_2\rangle + |l_2l_1\rangle \right)  
    \nonumber \\
    |\chi_2\rangle = \frac{1}{\sqrt{2}}\left(|l_1l_2\rangle - |l_2l_1\rangle \right)  
    \nonumber \\
    |\chi_3\rangle = \frac{1}{\sqrt{2}}\left(|l_1l_1\rangle + |l_2l_2\rangle \right) 
    \nonumber \\
    |\chi_4\rangle = \frac{1}{\sqrt{2}}\left(|l_1l_1\rangle - |l_2l_2\rangle \right)   
    \nonumber 
\end{split}
\end{equation}
the Hamiltonian becomes 

\begin{equation}
 H = \frac{1}{4}\left( 
\begin{array}{cc}
H^{(1)}& V^T \\
V & H^{hex} \\
\end{array} \right)
\end{equation}


\begin{equation}
{\renewcommand{\arraystretch}{1.5} 
\resizebox{0.9\textwidth}{!}{$
 H^{(1)} =\left( 
    \begin{array}{cccccc}
        2\Omega_1 & 0 & 0 & J  &J\sqrt{2}\cos\alpha & 0 \\
        0 & 2\Omega_2 & J & 0 & 0 & J\sqrt{2}\cos\beta  \\
        0 & J & -2\Omega_2 & 0 & 0 & -J\sqrt{2}\cos\beta \\
        J & 0 & 0 & -2\Omega_1 & -J\sqrt{2}\cos\alpha &  \\ 
        J\sqrt{2}\cos\alpha & 0 & 0 & -J\sqrt{2}\cos\alpha & -J\cos2\alpha & \frac{4B_z}{\cos\theta_2}-\frac{4B_z}{\cos\theta_1} \\
        0 & J\sqrt{2}\cos\beta &  -J\sqrt{2}\cos\beta & 0 & \frac{4B_z}{\cos\theta_2}-\frac{4B_z}{\cos\theta_1} & -J\cos2\beta  \\ 
    \end{array}
    \right)
$}}
\end{equation}

\begin{equation}
    H^{hex}=\left(\begin{array}{cc}
         J\cos2\beta & -\left(\frac{4B_z}{\cos\theta_2}+\frac{4B_z}{\cos\theta_1}\right) \\
         -\left(\frac{4B_z}{\cos\theta_2}+\frac{4B_z}{\cos\theta_1}\right) & J\cos2\alpha  \\
    \end{array}
    \right)
\end{equation}

\begin{equation}
V=\left( 
    \begin{array}{cccccc}
        0 & J\sqrt{2}\sin\beta & -J\sqrt{2}\sin\beta & 0 & 0 & 0   \\
        J\sqrt{2}\sin\alpha & 0 & 0 & -J\sqrt{2}\sin\alpha & 0 & 0  \\
    \end{array}
    \right)
\end{equation}

where $\alpha=(\theta_1+\theta_2)/2$ and $\beta=(\theta_1-\theta_2)/2$. Since usually $B_x^{(1)}, B_x^{(2)}\ll B_z$ it turns out that $\theta_1, \theta_2\ll 1$. Two limiting cases arise: (i) $B_x^{(1)} \approx B_x^{(2)}$ or (ii) $B_x^{(2)} \ll B_x^{(1)}$; (i) leads to $\cos\alpha=\cos\theta_1\approx 1-B_x^{(1)2}/8B_z^2$ and $\cos\beta=1$; (ii) gives $\cos\alpha \approx\cos\beta \approx 1+(B_x^{(1)2})/(4B_z)^2$. In both cases, one can take $\cos\alpha=\cos\beta=1$ to approximate $H$ by the upper left $6\times 6$ block (due $\sin\alpha=\sin\beta=0$ in the other blocks) as  
\begin{equation}
    H = \frac{1}{4}\left( 
    \begin{array}{cccccc}
        2\Omega_0 & 0 & 0 & J & \sqrt{2}J & 0  \\
        0 & 2\Omega_1 & J & 0 & 0 & \sqrt{2}J  \\
        0 & J & -2\Omega_1 & 0 & 0 & -\sqrt{2}J \\
        J & 0 & 0 & -2\Omega_0 & -\sqrt{2}J & 0  \\ 
        \sqrt{2}J & 0 & 0 & -\sqrt{2}J & -J & 0 \\
        0 & \sqrt{2}J &  -\sqrt{2}J & 0 & 0 & -J 
    \end{array}
    \right)
\end{equation}
with the leakage states given by 
\begin{equation}
\begin{split}
|\chi_+\rangle &= \frac{1}{\sqrt{2}}\left(|l_1l_2\rangle + |l_2l_1\rangle \right) \approx |T^0_{ba}\rangle \nonumber \\
|\chi_-\rangle &= \frac{1}{\sqrt{2}}\left(|l_1l_2\rangle - |l_2l_1\rangle \right) \approx |S_{ba}\rangle \nonumber .
\end{split}
\end{equation}
\section{Frequency dependence of leakage}
In the basis as $\{|00\rangle, |11\rangle, |\chi_1\rangle, |\chi_2\rangle, |01\rangle, |10\rangle, |\chi_3\rangle, |\chi_4\rangle \}$, the Hamiltonian with the exchange interaction becomes $H=\frac{\hbar}{4}H'$  with $H'$ :

\begin{equation}
\hspace*{1.5cm} 
{\renewcommand{\arraystretch}{1.5} 
\setlength{\arraycolsep}{1pt}   
\resizebox{0.8\textwidth}{!}{$
 \fl   H' = \left( 
    \begin{array}{cccccccc}
        2\Omega_1 & J & \sqrt{2}J\cos\alpha & 0 & 0 & 0 & J\sqrt{2}\sin\alpha & 0 \\
        J & -2\Omega_1 & -\sqrt{2}J\cos\alpha & 0 & 0 & 0 & -J\sqrt{2}\sin\alpha & 0 \\ 
        \sqrt{2}J\cos\alpha & -\sqrt{2}J\cos\alpha & -J\cos 2\alpha & \Delta\Omega & 0 & 0 & 0 & 0\\ 
        0 & 0 & \Delta\Omega & -J\cos 2\beta & J\sqrt{2}\cos \beta & -J\sqrt{2}\cos \beta & 0 & 0 \\ 
        0 & 0 & 0 & J\sqrt{2}\cos \beta & 2\Omega_2 & J & 0 & J\sqrt{2}\sin\beta\\
        0 & 0 & 0 & -J\sqrt{2}\cos\beta & J & -2\Omega_2 & 0 & -J\sqrt{2}\sin\beta \\ \hline
        J\sqrt{2}\sin\alpha & -J\sqrt{2}\sin\alpha & 0 & 0 & 0 & 0 & J\cos2\alpha & -8 B_z \\
        0 & 0 & 0 & 0 & J\sqrt{2}\sin\beta & -J\sqrt{2}\sin\beta & -8 B_z & J\cos2\beta  \\
    \end{array}
    \right)
$}}
\end{equation}

where $\Omega_1$ and $\Omega_2$ are the Bohr frequency of each qubit, $J=J_0+J_1(t)$ with $J_1(t)=J_1 (\sin\omega_1 t +\sin\omega_2 t)$. 
\begin{equation}
    \Delta B=4B_z \left(\frac{1}{\cos\theta_1}-\frac{1}{\cos\theta_2} \right) = 2(B_1 -B_2)
\end{equation}
accounts for the difference of moduli $B$ due to variations in the spin quantization axis between quantum dots.

The angles are $\alpha=(\theta_1+\theta_2)/2$, $\beta=(\theta_1-\theta_2)/2$, where $\tan \theta_i=B_x^{(i)}/2B_z$. Experimentally, $B_x^{(i)}\lesssim 100$ mT, and $B_z\gtrsim 600$ mT; hence $\tan \theta_i\lesssim 0.1$. Therefore 

\begin{align}
    \sin\theta_i/2 &\approx \frac{1}{2} \tan\theta_i = \frac{B^{(i)}_x}{4B_z}, \\
    \cos\theta_i/2 &\approx 1 -\frac{1}{8} \tan^2\theta_i = 1 -\frac{B^{(i)2}_x}{32B_z^2}, \\
    \cos\theta_i &\approx 1 - \frac{B^{(i)2}_x}{8 B_z^2}, \\
    \Delta B &\approx \frac{B^{(1)2}_x-B^{(2)2}_x}{2 B_z}.
\end{align}

Then, up to $O(B_x^2/B_z^2)$,
\begin{align}
    \sin\alpha &\approx \frac{B_x^{(1)}+B_x^{(2)}}{4B_z} = \frac{\Omega_1}{4\omega_z}\\
    \cos\alpha &\approx 1-\frac{B_x^{(1)}B_x^{(2)}}{16B_z^2} = 1-\frac{\omega_x^{(1)}\omega_x^{(2)}}{16\omega_z^2} \\
    \sin 2\alpha &\approx \frac{\Omega_1}{2\omega_z} \\
    \sin\beta &\approx \frac{B_x^{(1)}-B_x^{(2)}}{4B_z} = \frac{\Omega_2}{4\omega_z}\\
    \cos\beta &\approx 1+\frac{B_x^{(1)}B_x^{(2)}}{16B_z^2} = 1+\frac{\omega_x^{(1)}\omega_x^{(2)}}{16\omega_z^2} \\
    \sin 2\beta &\approx \frac{\Omega_2}{2\omega_z} \\
    \cos 2\alpha &= \cos^2\alpha-\sin^2\alpha \approx 1- \frac{\omega_x^{(1)}\omega_x^{(2)}}{8\omega_z^2} -\frac{\Omega_1^2}{16\omega_z^2}\\
    \cos 2\beta &= \cos^2\beta-\sin^2\beta \approx 1 +\frac{\omega_x^{(1)}\omega_x^{(2)}}{8\omega_z^2} - \frac{\Omega_2^2}{16\omega_z^2}
\end{align}
It can be seen that, for typical values of magntic fields (or frequencies), second order terms can be neglected so that cosine functions can be approximated by 1, and sine functions taken up to first orden in $B_x/B_z$. With this approximations,

\begin{equation}
\hspace*{1.5cm} 
{\renewcommand{\arraystretch}{1.5} 
\setlength{\arraycolsep}{2pt}   
\resizebox{0.8\textwidth}{!}{$
\fl    H = \frac{\hbar}{4}\left( 
    \begin{array}{ccc|ccc|cc}
        2\Omega_1 & J & \sqrt{2}J & 0 & 0 & 0 & J\sqrt{2}\frac{\Omega_1}{4\omega_z} & 0 \\
        J & -2\Omega_1 & -\sqrt{2}J & 0 & 0 & 0 & -J\sqrt{2}\frac{\Omega_1}{4\omega_z} & 0 \\ 
        \sqrt{2}J & -\sqrt{2}J & -J & \Delta \Omega & 0 & 0 & 0 & 0\\ \hline
        0 & 0 & \Delta \Omega & -J & J\sqrt{2} & -J\sqrt{2} & 0 & 0 \\ 
        0 & 0 & 0 & J\sqrt{2} & 2\Omega_2 & J & 0 & J\sqrt{2}\frac{\Omega_2}{4\omega_z}\\
        0 & 0 & 0 & -J\sqrt{2} & J & -2\Omega_2 & 0 & -J\sqrt{2}\frac{\Omega_2}{4\omega_z} \\ \hline
        J\sqrt{2}\frac{\Omega_1}{4\omega_z} & -J\sqrt{2}\frac{\Omega_1}{4\omega_z} & 0 & 0 & 0 & 0 & J & -8 \omega_z \\
        0 & 0 & 0 & 0 & J\sqrt{2}\frac{\Omega_2}{4\omega_z} & -J\sqrt{2}\frac{\Omega_2}{4\omega_z} & -8 \omega_z & J  \\
    \end{array}
    \right)
$}}
\end{equation}

If $J_0=J_1=0$ (non interacting qubits), the computational space is diagonal $H={\rm diag}(\pm\Omega_1/2,\pm\Omega_2/2)$, and the leakage space have energies $\pm \Delta B/4$ and $\pm 2B_z$.
\begin{equation}
    H = \frac{\hbar}{4}\left( 
    \begin{array}{cc|cc|cc|cc}
        2\Omega_1 & 0 & 0 & 0 & 0 & 0 & 0 & 0 \\
        0 & -2\Omega_1 & 0 & 0 & 0 & 0 & 0 & 0 \\ \hline
        0 & 0 & 0 & \Delta \Omega & 0 & 0 & 0 & 0\\ 
        0 & 0 & \Delta \Omega & 0 & 0 & 0 & 0 & 0 \\ \hline
        0 & 0 & 0 & 0 & 2\Omega_2 & 0 & 0 & 0\\
        0 & 0 & 0 & 0 & 0 & -2\Omega_2 & 0 & 0 \\ \hline
        0 & 0 & 0 & 0 & 0 & 0 & 0 & -8 \omega_z \\
        0 & 0 & 0 & 0 & 0 & 0 & -8 \omega_z & 0  \\
    \end{array}
    \right)
\end{equation}
Since $B_z\sim 10$ GHz, $\Omega_i\sim 1$ GHz, $J_0\sim 0.1$ GHz and $J_1\sim 0.01$ GHz, the approximate effect of $J_0$ is to shift the levels 
\begin{align}
    \Omega_i/2 &\rightarrow \sqrt{(\Omega_1/2)^2+(J_0/4)^2} \approx \frac{\Omega_i}{2} (1+\frac{J_0^2}{8\Omega_i^2}) \approx\Omega_i/2, \nonumber \\
    \pm \Delta B &\rightarrow J_0 \pm \Delta B, \nonumber \\
    \pm 2B_z &\rightarrow J_0 \pm 2B_z \approx \pm 2B_z, \nonumber
\end{align}
and to hybridize them, while $J_1(t)$ induce transitions between them.  
Then, the transitions to leakage states will occur at frequencies
\begin{equation}
    \omega = \frac{\Omega_i}{2} + \frac{J_0}{4} \pm \frac{B_x^{(1)2}-B_x^{(2)2}}{8B_z}.
\end{equation}

For instance, for $B_z=1$ T = 14 GHz, $B_x^{(1)} = 225,71$ mT = 3.16 GHz, $B_x^{(2)} = 60$ mT = 0.84 MHz, the shift of the leakage levels becomes $\Delta B/4 = 5.9$ mT = 82 MHz.


\end{document}